\documentclass{ieeeaccess}
\newcommand{\headerlogo}[1]{}
\newcommand{\headerlogoall}[1]{}
\usepackage[T1]{fontenc}
\usepackage{mathptmx}
\usepackage{cite}
\usepackage{amsmath,amssymb,amsfonts}
\usepackage{algorithmic}
\usepackage{graphicx}
\usepackage{textcomp}
\usepackage{caption}
\usepackage{subfig}

\usepackage{listings}
\usepackage{amsthm}
\usepackage{fvextra}
\newtheorem{lemma}{Lemma}
\def\BibTeX{{\rm B\kern-.05em{\sc i\kern-.025em b}\kern-.08em
    T\kern-.1667em\lower.7ex\hbox{E}\kern-.125emX}}
\newcommand{\EOD}{}

\begin{document}
\history{Date of publication xxxx 00, 0000, date of current version xxxx 00, 0000.}
\doi{10.1109/ACCESS.2017.DOI}

\title{Dynamic Strategy Adaptation in Multi-Agent Environments with Large Language Models}

\author{%
\uppercase{Shaurya Mallampati}\authorrefmark{1}\, (ORCID: 0009-0005-4079-9974),\,
\uppercase{Rashed Shelim}\authorrefmark{2,3}\, (ORCID: 0000-0002-6874-2110),\,
\uppercase{Walid Saad}\authorrefmark{2}\, (ORCID: 0000-0003-2247-2458),\,
\uppercase{and Naren Ramakrishnan}\authorrefmark{3}\, (ORCID: 0000-0002-1821-9743)%
}

\address[1]{Marriotts Ridge High School, Marriottsville, MD 21104, USA (e-mail: shauryasai@gmail.com)}
\address[2]{Department of Electrical and Computer Engineering, Virginia Tech, Alexandria, VA 22305, USA (e-mail: rasheds@vt.edu; walids@vt.edu)}
\address[3]{Department of Computer Science, Virginia Tech, Alexandria, VA 22305, USA (e-mail: naren@cs.vt.edu)}

\markboth
{Mallampati \headeretal: Dynamic Strategy Adaptation in Multi-Agent Environments with Large Language Models}
{Mallampati \headeretal: Dynamic Strategy Adaptation in Multi-Agent Environments with Large Language Models}

\corresp{Corresponding author: Shaurya Mallampati (e-mail: shauryasai@gmail.com).}

\begin{abstract}
Large language models (LLMs) demonstrate strong reasoning abilities across mathematical, strategic, and linguistic tasks. However, little is known about how well they reason in dynamic, real-time, multi-agent scenarios, such as collaborative environments where agents continuously adapt to each other’s behavior, as in cooperative gameplay settings. In this paper, we bridge this gap by introducing a novel framework that integrates LLM-based symbolic evaluation with reinforcement learning for real-time adaptation in cooperative, multi-agent environments without modifying the LLM weights or retraining during execution. Grounded in game-theoretic principles such as belief consistency and equilibrium stability, the framework enables agents to refine strategies during execution without retraining or centralized coordination. Our approach supports dynamic scenarios in which agents coordinate, communicate, and make decisions under continuously changing conditions. We provide real-time strategy refinement and adaptive feedback mechanisms that guide agents toward stable joint policies that approximate Nash equilibria, achieved empirically through iterative feedback and symbolic reward shaping. In contrast to previous efforts that evaluate LLM capabilities in static or turn-based settings, our method supports continuous mid-episode revision. Empirical results show that, in the most challenging combined-noise regime, our framework recovers approximately one reward unit relative to PPO-only baselines (about 2.6\% of the penalty floor), while maintaining decision latency around 1.6 milliseconds per step with less than a 2\% overhead compared to PPO. These findings demonstrate that integrating LLM-based symbolic evaluations with reinforcement learning enables more resilient, interpretable, and flexible multi-agent systems capable of sustaining robust collaboration in dynamic settings.
\end{abstract}

\begin{keywords}
Game Theory, Large Language Models, Multi-Agent Reinforcement Learning, Real-Time Strategy Adaptation, Reward Shaping, Symbolic Feedback
\end{keywords}

\setlength{\titlepgskip}{-15pt}

\maketitle

\section{Introduction}
\label{sec:introduction}
\PARstart{L}{arge} language models (LLMs) are an effective machine learning tool that can perform a broad range of reasoning tasks, such as mathematical problem solving \cite{ReActPrompting2023}, code generation \cite{SWEAgent2023,MetaGPT2023}, commonsense inference \cite{MindAgent2024,ThoughtAgents2024}, and strategic board games \cite{CiceroDiplomacy,GTBench}. However, their ability to operate effectively in real-time, multi-agent coordination settings remains under-explored~\cite{han2012game}. Given the rise of multi-agent systems, characterized by dynamic collaboration and automation, there is a need for new frameworks capable of supporting real-time strategic adaptation. Although game theory offers a robust mathematical framework for modeling strategic interactions, it typically assumes fixed decision rules that lack the flexibility to accommodate real-world complexity and change~\cite{han2012game}. In contrast, reinforcement learning (RL) provides a flexible mechanism for learning through experience. However, it requires extensive training and often produces strategies that lack clear interpretability. To guide this investigation, we ask the following research question: \textit{How can LLMs dynamically adapt and refine their game strategies based on real-time feedback in multi-agent, multi-turn environments to achieve enhanced rationality and equilibrium outcomes?}

A number of prior works have studied negotiation, tool integration, real-time planning, and embodied reasoning in the context of LLM-based decision-making~\cite{HeNegotiation2018, LewisBargaining2017, Toolformer2023, PaLME2023, ReAct2022, NegotiationArena2022, StarCraft2023, CiceroDiplomacy, CAMEL2023, AutoGen2023, LangAgent2023, Voyager}. For example, He et al.~\cite{HeNegotiation2018} and Lewis et al.~\cite{LewisBargaining2017} explore end-to-end learning through LLMs for negotiation dialogues, while Toolformer~\cite{Toolformer2023} demonstrates how LLMs can autonomously learn to invoke external tools during generation. PaLM-E~\cite{PaLME2023} highlights the value of integrating visual and proprioceptive inputs for grounded multimodal reasoning. ReAct prompting~\cite{ReAct2022} illustrates how models can reason and act by interleaving tool calls with intermediate steps. NegotiationArena~\cite{NegotiationArena2022} and StarCraft II benchmarks~\cite{StarCraft2023} provide multi-agent coordination evaluations, and CICERO~\cite{CiceroDiplomacy} showcases strategic interaction through dialogue in high-stakes, turn-based environments. Recent frameworks such as CAMEL~\cite{CAMEL2023}, AutoGen~\cite{AutoGen2023}, LangAgent~\cite{LangAgent2023}, and Voyager~\cite{Voyager} have introduced collaborative LLM teams with structured roles and planning objectives.
Other solutions such as SWE-agent~\cite{SWEAgent2023} and MetaGPT~\cite{MetaGPT2023} propose programming-based role assignment and skill composition pipelines, yet often operate with fixed dialogue states and precomputed coordination graphs. OctoAgents~\cite{OctoAgents2024} explore zero-shot LLM team formation but require architectural homogeneity and lack continual learning mechanisms. The MindAct~\cite{MindAct2024} and CoLLM~\cite{CoLLM2023} models examine collaboration over temporal chains but still assume a known world model. While prior solutions~\cite{HeNegotiation2018, LewisBargaining2017, Toolformer2023, PaLME2023, ReAct2022, NegotiationArena2022, StarCraft2023, CiceroDiplomacy, CAMEL2023, AutoGen2023, LangAgent2023, Voyager} enable symbolic reasoning, tool usage, and basic interagent dialogue, they typically do not support continuous strategic adaptation in cooperative, real-time environments. However, most existing approaches operate in static settings or rely on single-shot policy inference \cite{GTBench,GameTheoryLLM,RationalLLM,Toolformer2023,ReAct2022}. There is little ability for agents to revise goals, respond to degraded teammate performance, or restructure plans in response to unforeseen environment transitions. A central open challenge is, thus, how to enable agents to adjust their strategy mid-execution using context-aware feedback, memory persistence, symbolic priors, and feedback-aware control loops that reflect real-time game-theoretic dynamics.

The strategic use of LLMs in dynamic multi-agent environments face significant challenges in enabling agents to reason adaptively, coordinate under
uncertainty, and revise decisions as environments evolve. While early systems~\cite{GTBench, GameTheoryLLM, RationalLLM, Voyager, EmergentComm1, EmergentComm2, CiceroDiplomacy} have made progress in specific areas, such as static equilibrium modeling, symbolic dialogue protocols, or role-based planning, they remain limited in three key aspects: 1) they cannot revise their beliefs which means they cannot update internal models of the environment or other agents’ intentions when goals shift or conditions become noisy, 2) they lack real-time mechanisms for coordination across agents, and 3) they do not incorporate symbolic priors (structured knowledge constraints) or game-theoretic principles (strategic reasoning frameworks such as equilibrium or bargaining models) into feedback loops. These limitations highlight the need for integrated multi-agent reasoning frameworks that support real-time strategy adaptation and structured coordination, guided by symbolic feedback and contextual understanding under uncertainty.

Although CICERO~\cite{CiceroDiplomacy} offers a rich symbolic policy in the context of Diplomacy, it is limited to turn-based play, where strategic reasoning and belief updates occur only at discrete turn boundaries, limiting responsiveness to real-time state changes and mid-episode coordination failures. Reinforcement learning from human feedback (RLHF)~\cite{RLHF} allows reward shaping but lacks dynamic revision based on environment evolution. Existing frameworks such as RolePlay~\cite{RolePlay2023} and ThoughtAgents~\cite{ThoughtAgents2024} aim to inject personality and dialogue abstraction but do not connect memory or belief state with performance feedback. Few systems track and utilize contextual priors across time steps. Even fewer are capable of aligning their behavioral revisions with task-level metrics like regret, coordination efficiency, or Nash deviation. We argue that bridging this gap requires dynamic feedback systems that are tightly coupled with symbolic modeling and game-theoretic planning principles. This naturally brings us back to the broader challenge of continuous strategic adaptation in cooperative, real-time environments.

Building on this motivation, we examine continuous strategic adaptation more directly. The key difficulty is enabling LLM-based agents to revise their strategies in real-time as goals shift, agents deviate, or coordination structures deteriorate. While prior work introduces symbolic reasoning or dialogue abstractions, these systems fall short of coupling behavioral revisions with long-horizon planning, leaving strategic adaptation incomplete in evolving environments.

Beyond continuous adaptation, agents must also cope with uncertainty, incomplete information, and evolving opponent behavior. In practice, they must not only adapt their strategies continuously, but also reason under partial observability and dynamic conditions. An examination of the literature that addresses uncertainty, partial observability, and strategic adaptation in games reveals several key limitations~\cite{GTBench, GameTheoryLLM, RationalLLM, Voyager, EmergentComm1, EmergentComm2, CiceroDiplomacy}. GTBench~\cite{GTBench} tests LLMs on game-theoretic matrix tasks, finding poor robustness in partially observable environments. GameTheoryLLM~\cite{GameTheoryLLM} formalizes Bayesian agent workflows for negotiation but imposes synchronous update protocols. RationalLLM~\cite{RationalLLM} studies equilibrium behavior but does not support belief evolution or policy trajectory tracking. Multi-Agent Reinforcement Learning benchmarks like MAgent~\cite{MAgent2018} and EnvAgent~\cite{EnvAgent2023} test emergent roles but fail to align agents with structured symbolic constraints. Systems like MindAgent~\cite{MindAgent2024}, DEAM~\cite{DEAM2024}, and ToM-GPT~\cite{ToMGPT2024} simulate theory-of-mind and belief graphs but rely on heavily curated scenarios. CoaLLM~\cite{CoaLLM2023} and OpenAgents~\cite{OpenAgents2023} attempt to unify self-play and LLM planning, though they still assume pre-scripted task flows. Taken together, these approaches demonstrate progress in modeling interaction and adaptation, yet they remain fragmented and constrained by narrow experimental setups. None provide a unified framework for handling uncertainty, partial observability, and continuous adaptation in open-ended, real-time environments.

The main contribution of this paper is to address the limitations of static decision-making methods, including rule-based strategies, turn-based reasoning, and offline reinforcement learning, by proposing a novel framework that enables agents to adapt strategies in real time using LLM-guided reward shaping. By integrating symbolic evaluations from a frozen language model with a reinforcement learning loop, the proposed system refines agent behavior based on contextual signals such as order progress (the evolving state of tasks or objectives being completed within the environment) and timestep. This hybrid framework is designed to support collaboration in unpredictable, noisy environments where conventional policies tend to fail. In addition to developing the architecture, the proposed framework is empirically evaluated across dynamic environments to assess performance, robustness to noise, and latency tradeoffs. The proposed method delivers strategic benefits while preserving real-time inference speeds. In summary, the key contributions of this paper are:

\begin{itemize}\setlength{\itemsep}{0.3em}
\item We evaluate our framework within the Overcooked-AI~\cite{wang2020overcooked} simulation environment. 
Across four dynamic regimes (i.e., No Noise, Noise, Delay, and Combo), we develop a real-time shaping system that connects 
proximal policy optimization (PPO) agents with frozen LLM evaluations. Prompt-based symbolic feedback allows agents to 
revise decisions mid-episode in response to uncertainty, teammate variability, and timing disruptions. Agents produce 
structured prompts encoding state, goals, and short-term history, and integrate the returned judgments into a 
latency-constrained reward-shaping loop.

  \item We formalize symbolic feedback in multi-agent coordination by adapting game-theoretic principles such as belief revision, strategic consistency, and role alignment into a reinforcement learning-compatible reward-shaping pipeline. This symbolic layer guides agents toward cooperative behaviors by enforcing priors on mutual belief, future intent, and role commitment, without requiring full observability or centralized policy access.
\item Across four randomized, high-noise Overcooked-AI regimes (Noise, Delay, and Combo, together with the clean No Noise reference), empirical results show that PPO+LLM provides modest but measurable robustness benefits. In the most challenging combined-noise setting, PPO+LLM recovers approximately one reward unit relative to PPO-only baselines (about 2.6\% of the 40-point penalty floor), indicating improved stability under simultaneous observation noise and timing penalties. These gains are consistent across seeds and reflect the framework’s ability to preserve cooperative task flow under perturbations.

\item PPO+LLM agents maintained real-time suitability, operating at approximately 1.6 milliseconds per decision step with less than a 2\% overhead relative to PPO. Despite the added symbolic evaluation, latency remained within the same narrow range observed for all baselines. While the sparse-reward configuration does not yield full-task completions, PPO+LLM showed slightly improved coordination under combined disturbances, reflected by small but consistent gains in recovered reward relative to the penalty floor. Together, these results indicate that symbolic shaping preserves real-time performance while providing modest robustness benefits in the most challenging regimes.

 \item Finally, in the dense-reward Burger Kitchen scenario, PPO+LLM agents tended to converge toward Nash-like joint policies, where unilateral deviations offered no advantage. These stable cooperative behaviors emerged from symbolic feedback alone, without joint training, parameter sharing, or centralized supervision.

\end{itemize}
The remainder of this paper is organized as follows.
Section~\ref{sec:architecture} presents the system architecture, including the real-time symbolic shaping
pipeline, the LLM-based feedback mechanism, and its integration with PPO; it also defines the approximate
Nash-equilibrium probe used to evaluate stability under unilateral deviations.
Section~\ref{sec:results} reports simulation results across four Overcooked-AI regimes, covering performance
under noise, robustness to structured penalties, task completion behavior, latency characteristics, and
equilibrium behavior as measured by Nash gaps.
Section~\ref{sec:conclusion} concludes with a summary of findings, limitations, and directions for future work
on real-time strategic adaptation using language-based feedback.
Finally, the appendices provide additional implementation and reproducibility details, including model and
system complexity, wrapper design, and the best-response training and evaluation infrastructure
(Appendix~\ref{sec:complexity}).

\begin{figure}[t]
    \centering
    \includegraphics[width=\linewidth]{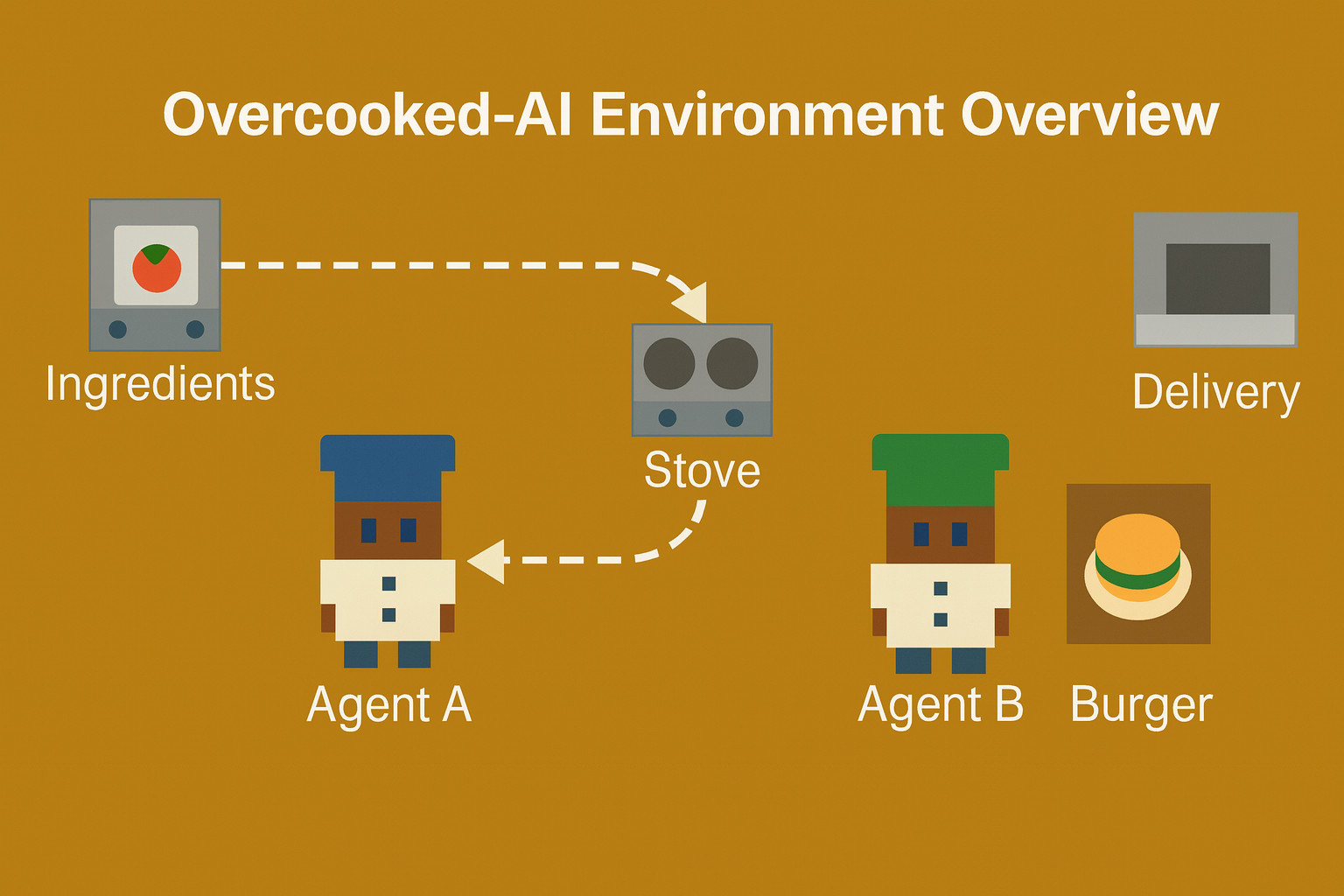}
    \caption{Overview of the Overcooked-AI cooperative environment. The environment is represented as a grid-based kitchen containing two agents, interactive objects (e.g., ingredient sources, cooking stations, assembly items), and task-relevant regions such as delivery counters. Agents must coordinate spatially and temporally to complete shared cooking tasks under movement and interaction constraints.}
    \label{fig:overcooked_ai_overview}
\end{figure}

\begin{figure}[t]
  \centering
  \includegraphics[width=\columnwidth]{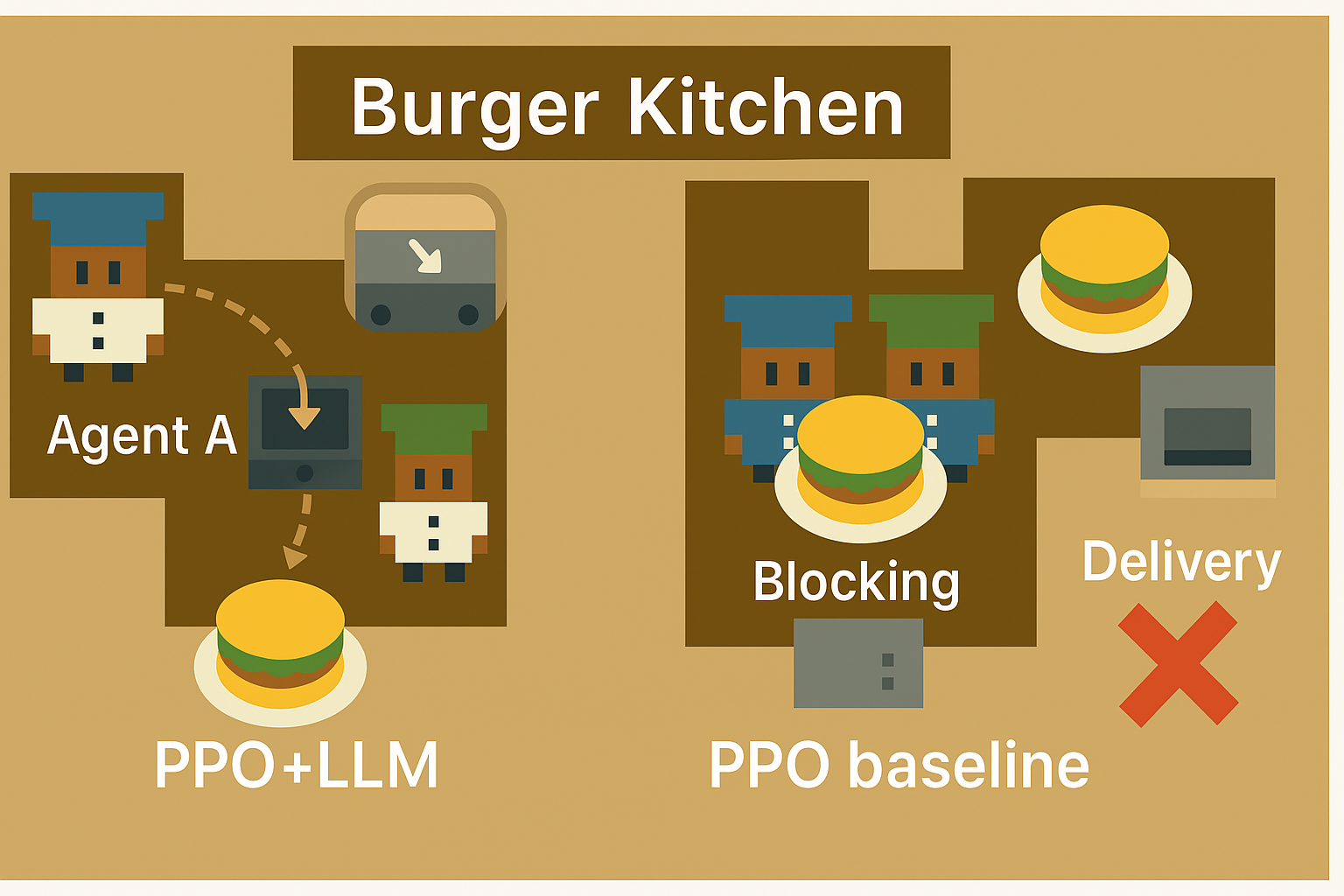}
  \caption{Custom Overcooked-AI "Burger Kitchen" scenario used in our experiments.
  Agent A (blue) retrieves raw patties from the fridge and cooks them on the \textit{Stove},
  while Agent B (green) assembles cooked patties with buns and condiments at the \textit{Burger Station}
  and delivers the finished burger through the \textit{Delivery Window}. Dashed arrows indicate task flow
  between stations. This scenario imposes strict temporal dependencies and spatial constraints that require
  real time role specialization, dynamic coordination, and mid episode strategy adjustment, exactly the
  challenges our PPO+LLM shaping framework is designed to address.}
  \label{fig:overcooked_ai}
\end{figure}

\begin{figure}[t]
\centering
\includegraphics[width=\columnwidth]{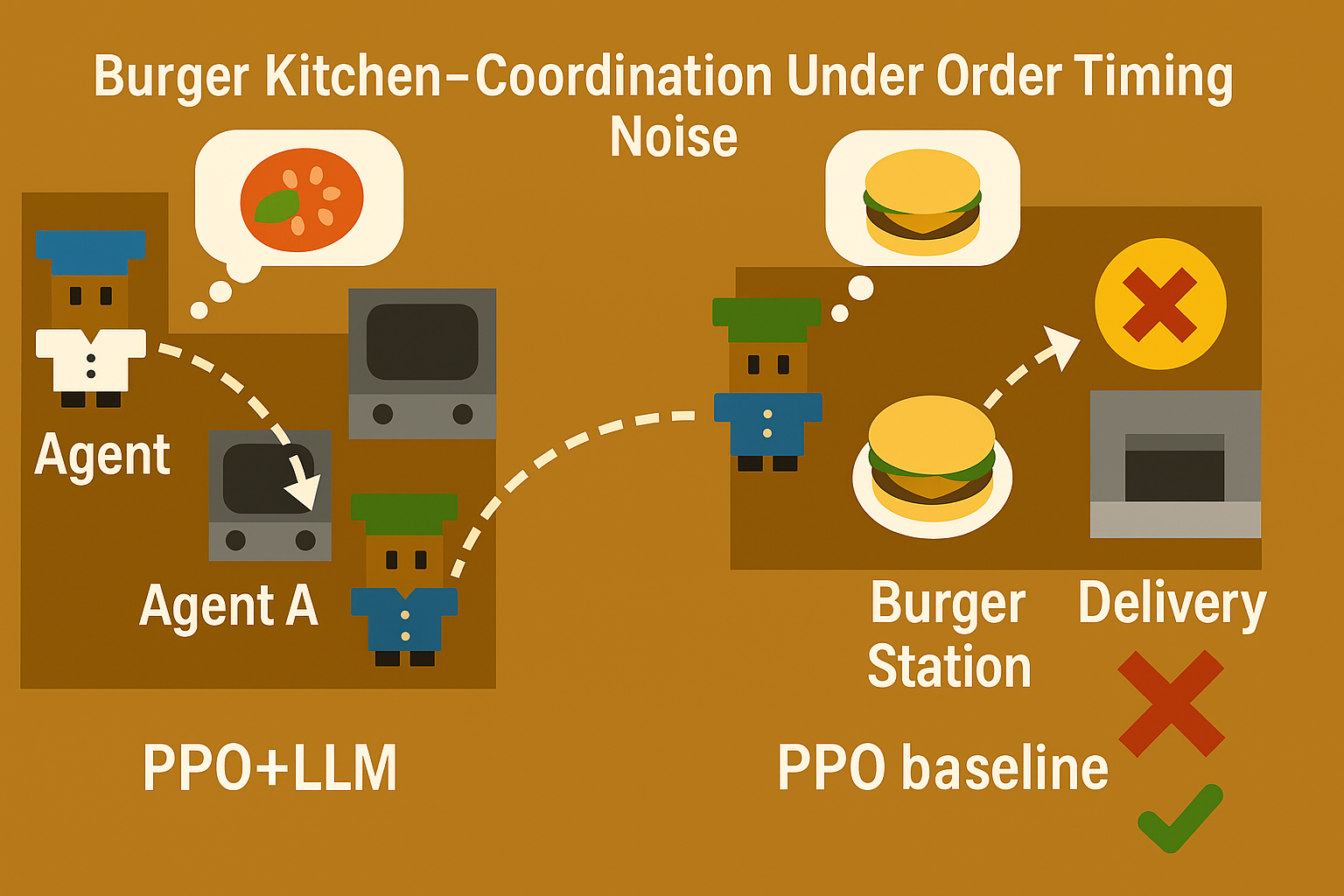}
\caption{\textbf{Burger Kitchen: Coordination Under Order Timing Noise.}
Left: PPO+LLM agents adapt to temporal perturbations and successfully coordinate task flow despite misaligned order signals.
The tomato icon represents an off-timing ingredient request, which the symbolic feedback loop helps agents recover from.
Right: PPO baseline agents fail to deliver the completed burger due to miscoordination, resulting in idle output and a missed delivery.
This figure illustrates how symbolic feedback enables robust strategy adaptation under timing noise.
Results averaged over $n = 200$ evaluation episodes.}
\label{fig:env_order_timing}
\end{figure}

\section{System Architecture}\label{sec:architecture}
\begin{figure*}[t!]
\vspace{-0.15in} 
  \centering
 \includegraphics[width=\textwidth]{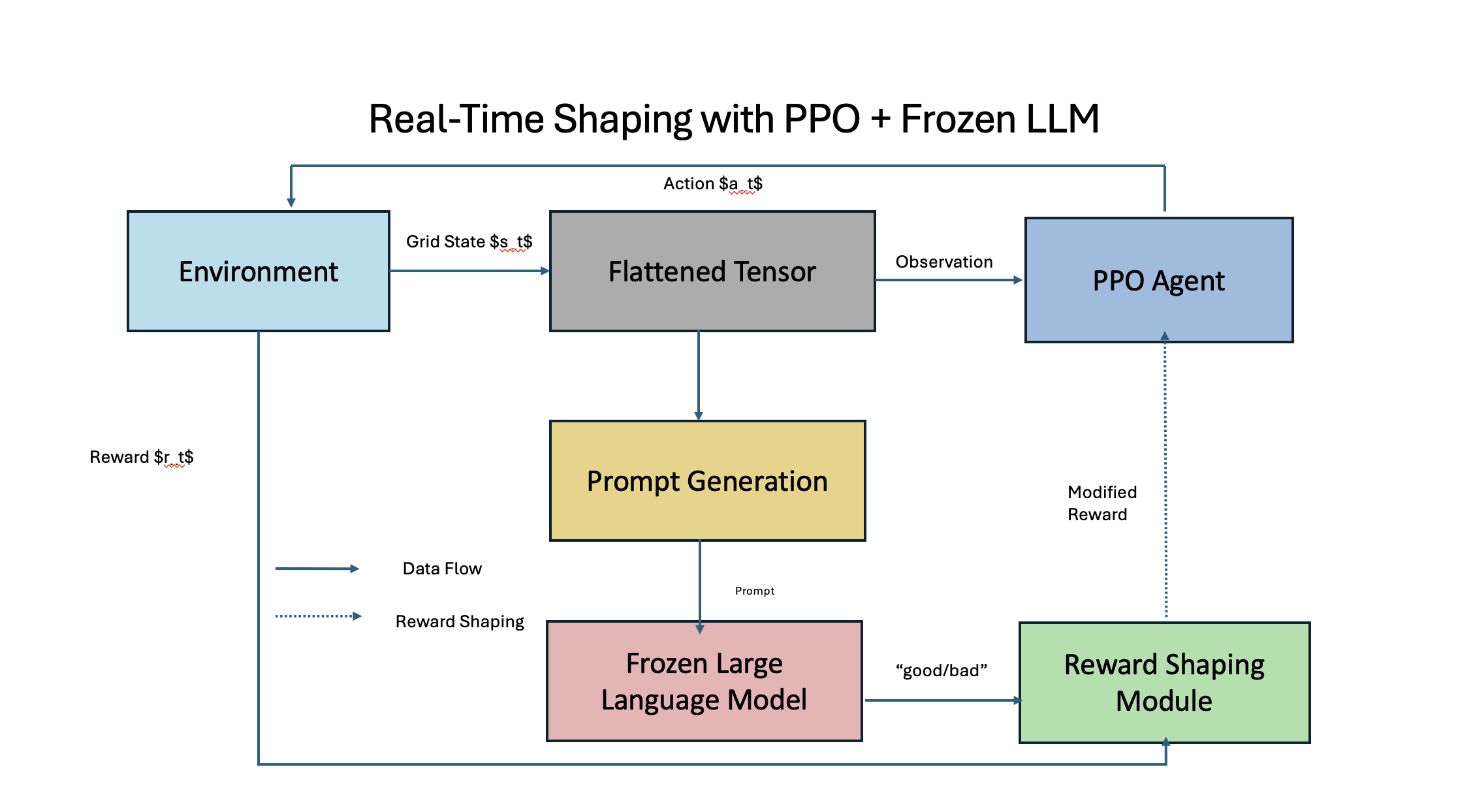}
  \caption{\textbf{Real-Time Shaping System with PPO + Frozen LLM.} The environment emits a grid-state at each timestep. That grid is \textit{flattened} into a 1D observation vector and passed to both the PPO agent and a prompt generator. In parallel, a \textit{prompt generator} encodes task context (e.g., remaining orders and step count) into a text prompt. This prompt is evaluated by a frozen LLM (GPT-Neo 1.3B), whose binary ``good''/``bad'' signal is mapped to a shaping bonus via the reward-shaping module. The PPO agent then updates its policy using this \textit{shaped} reward signal.}
\label{fig:architecture}

\vspace{-0.20in} 
\end{figure*}

\begin{figure}[t]
\centering
\includegraphics[width=\columnwidth]{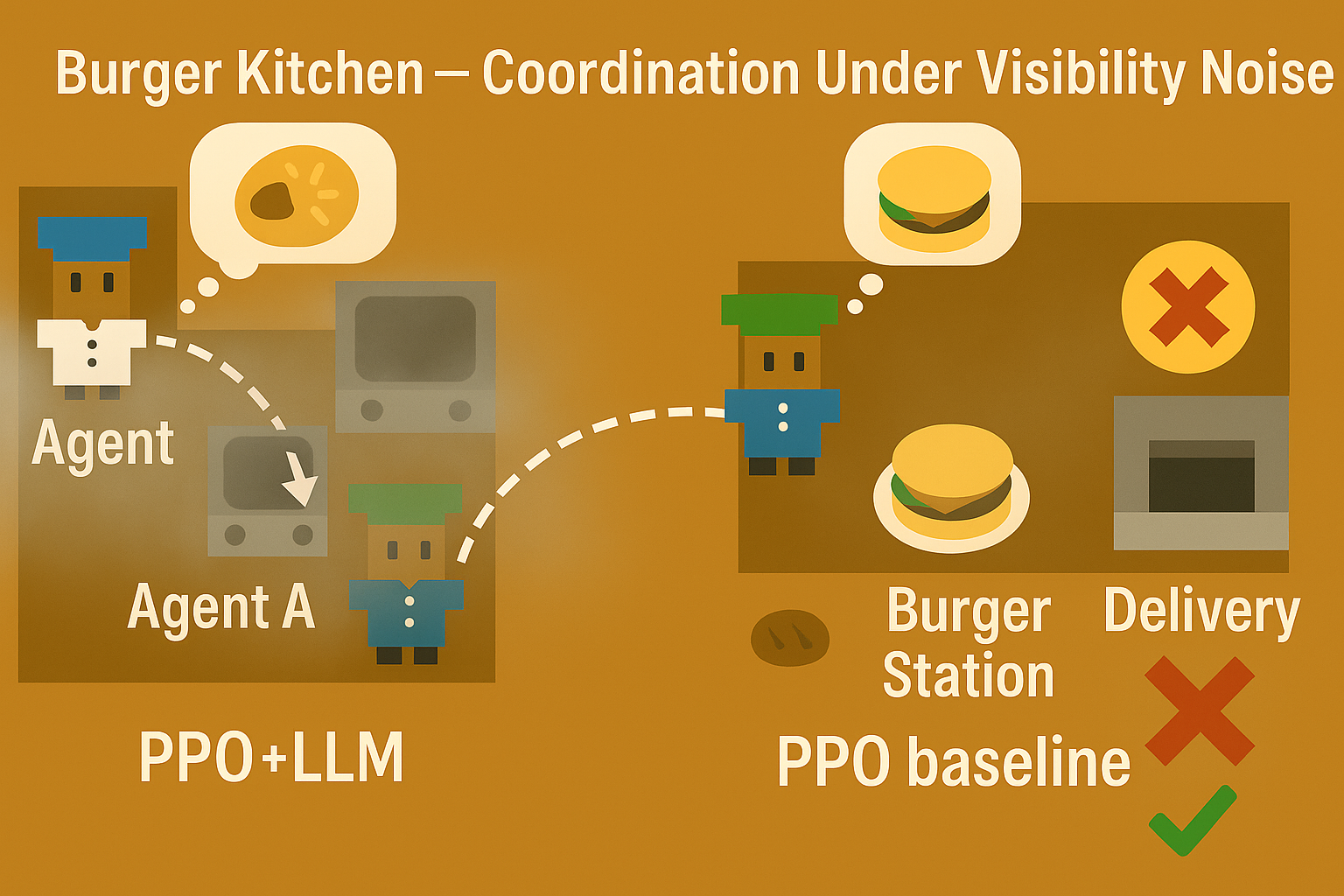}
\caption{\textbf{Burger Kitchen: Coordination Under Observation Noise.}
Left: PPO+LLM agents maintain effective collaboration despite partial visual occlusion.
Even when perceptual details are masked or distorted, symbolic feedback helps agents preserve task flow and avoid conflicts.
Right: PPO baseline agents fail to recognize delivery opportunities due to limited visibility, leading to idle burgers and missed outputs.
This figure demonstrates how language based shaping enables perceptual robustness and consistent goal pursuit under noisy observations.}
\label{fig:env_observation_noise}
\end{figure}

\begin{figure}[t]
\centering
\includegraphics[width=\columnwidth]{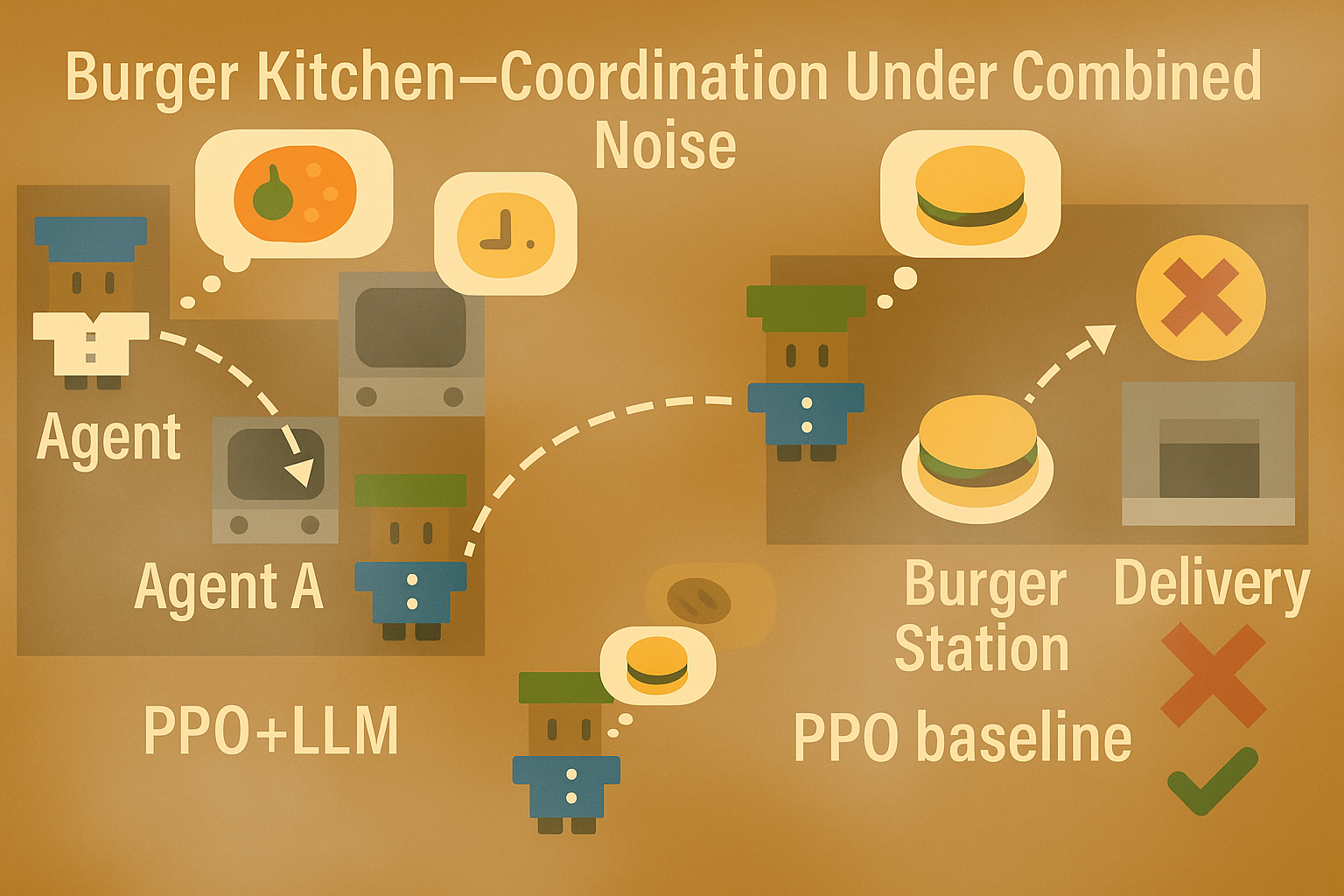}
\caption{\textbf{Burger Kitchen: Coordination Under Combined Noise.}
\textit{Left:} PPO+LLM agents overcome simultaneous order-timing jitter (clock icon) and
visibility corruption (fog and pixel-noise tile), adapting roles and completing delivery.
\textit{Right:} PPO baseline agents fail under the same disturbances. Perception noise
causes hesitation, timing jitter breaks the cook–deliver sequence, and the burger
idles at the window (×). Prompt-based feedback enables robustness where reactive
policies collapse.}
\label{fig:env_combined_noise}
\end{figure}

We instantiate our experimental environment using Overcooked-AI~\cite{wang2020overcooked}, a cooperative multi-agent simulator in which agents must coordinate to cook and deliver meals under time pressure and spatial constraints. At each time step \(t\), the environment emits a grid state \(s_t\) that captures the spatial layout of \(N=2\) agents, interactive objects (such as ingredients and plates), and task-relevant regions (such as delivery counters). Each grid cell encodes the type and state of any entity occupying it, and the grid is subsequently transformed into a one-dimensional observation vector to ensure compatibility with both the PPO policy network and the LLM prompt generator. Figure~\ref{fig:overcooked_ai_overview} illustrates our specific ``Burger Kitchen'' configuration, where one agent prepares cooked patties while the other assembles and delivers finished burgers. This layout introduces strict temporal dependencies and spatial bottlenecks that require dynamic role allocation and real-time strategic coordination.

Figure~\ref{fig:overcooked_ai_overview} provides a schematic overview of the Overcooked-AI
grid-world and the interactive stations that define the cooperative cooking task.
Figure~\ref{fig:overcooked_ai} then instantiates this template as our Burger Kitchen layout,
highlighting the role split between cooking (Agent A) and assembly plus delivery (Agent B),
and the resulting temporal dependencies that require coordination.

In this environment, the symbolic reward shaping framework is instantiated over spatially
separated roles. By linking symbolic judgment to policy adaptation, the framework gradually
enforces game-theoretic alignment. Both PPO agents respond independently to the same symbolic
signal and adapt their behavior to maximize cooperative feedback. Over time, they converge
toward a stable configuration in which each agent's role complements the other.

Figure~\ref{fig:architecture} provides an overview of the real-time shaping architecture
that integrates proximal policy optimization (PPO) with a frozen large language model (LLM).
At each timestep \(t\), the environment emits a grid-based state \(s_t\), which is flattened
into a one-dimensional observation vector. This observation is passed to the PPO policy for
action selection, while a parallel prompt generation module extracts high-level task context,
such as remaining orders and the current timestep. The resulting textual prompt is evaluated
by a frozen LLM, which produces a symbolic judgment of cooperative progress. This judgment is
mapped to a bounded shaping signal that augments the environment reward before policy updates.
By decoupling symbolic evaluation from policy learning, the architecture enables real-time
strategy adaptation without modifying the LLM weights or introducing centralized coordination.

Figure~\ref{fig:env_order_timing} illustrates a representative Burger Kitchen rollout under
order-timing perturbations. PPO+LLM maintains coherent task flow despite misaligned order
signals, while the PPO baseline frequently miscoordinates and fails to complete delivery,
demonstrating how symbolic shaping stabilizes cooperation when temporal cues are perturbed.

Beyond aggregate return statistics, we qualitatively examine how symbolic shaping
affects coordination behavior under different sources of uncertainty in the
Burger Kitchen environment. In particular, we consider settings in which agents
must operate under partial observability, temporal perturbations, or both.
These scenarios are designed to stress coordination by degrading the reliability
of low-level sensory cues and disrupting the temporal alignment between subtasks,
conditions under which purely reactive policies are prone to miscoordination.

We first analyze coordination under observation noise, where the agents’
feature-level observations are corrupted by additive Gaussian perturbations.
This setting preserves the underlying task structure but obscures fine-grained
perceptual details required for precise timing and spatial alignment. As shown
in Figure~\ref{fig:env_observation_noise}, PPO+LLM agents maintain coherent role
execution despite partial visual occlusion, continuing to advance the cook–assemble–deliver
pipeline. In contrast, PPO baseline agents frequently fail to recognize delivery
opportunities when perceptual cues are degraded, resulting in idle burgers and
missed task completion. This comparison illustrates that symbolic feedback can
provide a stabilizing high-level signal that compensates for unreliable sensory
input.

We next consider the most challenging regime, which combines observation noise
with stochastic order-timing perturbations. This setting simultaneously degrades
perception and disrupts the temporal dependencies between cooking and delivery,
making coordination failures especially likely. Figure~\ref{fig:env_combined_noise}
shows that PPO+LLM agents adapt their behavior to recover task flow under these
compound disturbances, preserving role specialization and completing delivery.
By contrast, PPO baseline agents often hesitate or break the cook–deliver sequence,
causing the finished burger to stall at the delivery window. These qualitative
examples reinforce the quantitative results by illustrating how symbolic shaping
supports robust coordination when both perceptual and temporal cues are unreliable.

This convergence reflects a form of equilibrium stability in the learned joint policy: once both agents settle into mutually compatible strategies, unilateral deviations no longer offer any benefit. Crucially, the framework does not compute an equilibrium in a closed-form or analytic sense. Instead, the stability arises organically through repeated interaction and the shaping feedback delivered by the LLM. The agents refine their behavior over time using only local observations and symbolic signals, without requiring access to one another’s internal policy state or any explicit coordination channel.

The PPO agent, implemented using Stable-Baselines3~\cite{raffin2021stable}, takes the observation \(o_t\) and computes an action \(a_t\) based on its learned policy \(\pi_\theta(a_t \mid o_t)\), where \(\theta\) represents the parameters of the neural network. The policy maps observed features to a discrete set of agent behaviors, such as movement or object interaction.

This action is then executed in the environment, which returns a scalar reward \( r_t \) and the next state \( s_{t+1} \). In parallel, a prompt generation module extracts key contextual signals, such as the number of remaining orders and the current timestep, and formats them into a text prompt \( \Phi(s_t) \) (e.g., \texttt{"orders:2 t:25/400"}).

This textual prompt is evaluated by a frozen large language model (GPT-Neo 1.3B). 
The LLM produces logits for specific tokens such as \texttt{"good"} and 
\texttt{"bad"}, which reflect the model’s confidence that the agents are 
cooperating effectively. If the logit for \texttt{"good"} exceeds that for 
\texttt{"bad"}, a scalar bonus \( \lambda_s \) is applied:

\vspace{-0.08in}
\begin{equation}
\lambda_s = 
\begin{cases}
\lambda_{s,\text{bonus}}, & \text{if } \text{logit}(\texttt{"good"}) > \text{logit}(\texttt{"bad"}) \\
0, & \text{otherwise}.
\end{cases}
\label{eq:llm_shaping}
\end{equation}
\vspace{-0.05in}

If the LLM provides a positive evaluation, meaning that the high-level assessment
of the current state--action context is favorable (i.e., approved rather than
rejected), an additional shaping bonus is applied to the environment reward.
Since this bonus directly affects the reward signal used for policy optimization,
it is necessary to ensure that the resulting shaped reward remains bounded in
order to preserve stable advantage estimation and PPO updates.

Using Eq.~(\ref{eq:shaped_reward}), the shaped reward is defined as
\[
r'_t = r_t + \lambda_s\, c_t,
\]
where \( c_t \in [-1,1] \) denotes a bounded symbolic shaping signal produced by
the frozen LLM. In our implementation, \(c_t\) is nonzero only when a positive
evaluation is returned.

The following lemma establishes that this shaped reward remains bounded.

\begin{lemma}[Bound on the Shaped Reward]
\label{lem:shaping_bound}
Assume the environment reward satisfies \( r_t \in [r_{\min},\, r_{\max}] \), and
the symbolic shaping signal satisfies \( c_t \in [-1,1] \), with shaping coefficient
\( \lambda_s \ge 0 \).
Then the shaped reward defined in Eq.~(\ref{eq:shaped_reward}),
\[
r'_t = r_t + \lambda_s\, c_t,
\]
is bounded for all \( t \) and satisfies
\[
r'_t \in [\, r_{\min} - \lambda_s,\; r_{\max} + \lambda_s \, ].
\]
In particular, the shaping term cannot cause unbounded reward growth or destabilize
advantage estimation.
\end{lemma}

\begin{proof}
Since \( r_t \in [r_{\min}, r_{\max}] \) and \( c_t \in [-1,1] \), we have
\[
r'_t = r_t + \lambda_s c_t \ge r_{\min} - \lambda_s,
\qquad
r'_t = r_t + \lambda_s c_t \le r_{\max} + \lambda_s.
\]
Thus \( r'_t \) is bounded in the stated interval.
\end{proof}

Otherwise, the original reward \( r_t \) is retained. The shaping coefficient \( \lambda_s \) is manually tuned to remain small and bounded, ensuring stability during PPO training. Finally, to guarantee that shaping never dominates the task reward, $\lambda_s$ must satisfy \begin{equation} \lambda_s \le \min\!\left\{ (1-\gamma)\,\delta_V, \; \tfrac{1-\gamma}{2}\,\varepsilon_{\text{clip}}\,A_{\text{scale}} \right\}. \label{eq:lambda_bound} \end{equation} where $\delta_V$ is the maximum tolerated critic-target drift, $\varepsilon_{\text{clip}}$ is the PPO clipping parameter, and $A_{\text{scale}}$ is a characteristic scale of the unshaped advantages. This bound ensures that shaping acts only as a small regularizer rather than overwhelming the base reward signal.

The shaped reward \( r'_t \) is used to compute the advantage function \( \hat{A}_t \), which guides PPO policy updates. We adopt generalized advantage estimation (GAE)~\cite{raffin2021stable}, which links short-term reward shaping with long-horizon credit assignment. Let \( \gamma \in [0,1) \) denote the standard discount factor governing the agent’s valuation of future rewards. The temporal-difference error is defined as
\begin{equation}
\delta_t = r'_t + \gamma V(s_{t+1}) - V(s_t),
\label{eq:td_error}
\end{equation}
where $s_t$ denotes the environment state at timestep $t$.

Using these temporal-difference errors, the generalized advantage estimator (GAE) is computed as
\begin{equation}
\hat{A}_t
= \delta_t
+ (\gamma \lambda_G)\,\delta_{t+1}
+ (\gamma \lambda_G)^2\,\delta_{t+2}
+ \cdots
+ (\gamma \lambda_G)^{\,T - t - 1}\,\delta_{T-1},
\label{eq:gae}
\end{equation}

Because the LLM remains frozen, its evaluations are deterministic and repeatable.
In our implementation, we bypass stochastic decoding entirely and instead read the
raw model logits for specific tokens (e.g., ``good'' vs.\ ``bad'') rather than
sampling. This ensures that the symbolic feedback signal is stable and reproducible
across identical prompts. To maintain training stability, the shaping coefficient
$\lambda_s$ is kept small and explicitly bounded, preventing destabilization or
runaway updates. Formally, if the shaped reward is defined as
\begin{equation}
r'_t = r_t + \lambda_s\, c_t,
\qquad c_t \in [-1,1],
\label{eq:shaped_reward}
\end{equation}
where $c_t$ denotes a bounded symbolic shaping signal produced by the frozen LLM
and is distinct from the environment state $s_t$, then the per-step perturbation
satisfies $|r'_t - r_t| \le \lambda_s$. Over the discounted horizon, and for a fixed
policy $\pi$, the cumulative shift in value estimates is bounded by

\begin{equation}
\big| V'^{\pi}(s) - V^{\pi}(s) \big|
\le \frac{\lambda_s}{1 - \gamma}.
\label{eq:value_bound}
\end{equation}

The corresponding shift in advantages is
\begin{equation}
\big| A'^{\pi}(s,a) - A^{\pi}(s,a) \big|
\le \frac{2\lambda_s}{1 - \gamma}.
\label{eq:advantage_bound}
\end{equation}

These bounds ensure that both value targets and advantages remain close to their original scale during PPO training.

Beyond stability, these evaluations instantiate two fundamental game-theoretic principles.
The first is \textit{belief revision}, which parallels Bayesian updating.

Recall that the LLM produces a discrete symbolic judgment (e.g., ``good'' vs.\ ``bad'') reflecting
whether the agents’ joint behavior appears cooperative.
We model this judgment using a latent binary variable
\[
z \in \{\text{cooperative},\;\text{non-cooperative}\}.
\]

Let $b_t \in [0,1]$ denote the LLM’s belief at time $t$ that the agents are cooperating,
conditioned on the textual prompt $\phi(s_t)$ extracted from the environment state.
This belief is updated using a Bayesian revision rule:
\begin{equation}
\label{eq:belief_revision}
b_{t+1}
=
\frac{
P(\phi(s_{t+1}) \mid z)\, b_t
}{
P(\phi(s_{t+1}) \mid z)\, b_t
+
P(\phi(s_{t+1}) \mid \neg z)\, (1 - b_t)
}.
\end{equation}

Here, $z$ denotes the latent event that the agents’ joint behavior is \emph{cooperative},
while $\neg z$ denotes \emph{non-cooperative} behavior.
Operationally, these cases correspond to the LLM assigning greater logit mass to positive
(cooperative) or negative (non-cooperative) symbolic evaluations, respectively.
Thus, the sign of the LLM’s internal evaluation implicitly determines whether the observed
prompt evidence supports cooperation or non-cooperation.

The quantity $b_{t+1}$ represents the LLM’s posterior belief at the next timestep that the
agents are cooperating, updated in response to the newly observed prompt derived from $s_{t+1}$.
This revision captures how symbolic feedback—such as stalled orders, role conflicts, or
inefficient task allocation—modulates cooperative belief over time.
Because the LLM is frozen and evaluated deterministically, $b_t$ evolves predictably without
requiring any retraining of the underlying policy.

The second principle is \textit{strategic consistency}, which is implemented by reinforcing only those actions that align with collaborative success. Formally, the shaping signal is applied as
\begin{equation}
c_t =
\begin{cases}
+1 & \text{if } a_t \in \mathcal{A}_{\text{coop}}, \\
0  & \text{otherwise},
\end{cases}
\qquad c_t \in [-1,1].
\label{eq:strategic_consistency}
\end{equation}

Here, $\mathcal{A}_{\text{coop}}$ denotes the subset of actions that the frozen LLM evaluates as contributing to cooperative task flow in state $s_t$. In practice, an action is included in $\mathcal{A}_{\text{coop}}$ when the LLM assigns higher logit mass to the token \texttt{"good"} than to \texttt{"bad"} for the prompt describing $(s_t, a_t)$. These are typically actions that progress the shared objective, maintain role specialization, prevent collisions, or advance pending orders. By restricting the shaping signal to this cooperative set, the framework rewards only behavior that meaningfully supports joint success, discourages unstable or selfish deviations, and strengthens long-horizon coordination.


Putting these components together, PPO+LLM embeds structured symbolic evaluation
directly into the reinforcement learning loop. At each timestep, a prompt
generation module encodes task-relevant context (e.g., pending orders, timestep,
and relative agent positions) into a fixed text template that is scored by a
frozen LLM using the logits of tokens such as ``good'' and ``bad.'' If the logit
for ``good'' exceeds that for ``bad,'' a shaping bonus $\lambda_s$ is applied as
in Eq.~\eqref{eq:llm_shaping}, yielding the shaped reward (Lemma~\ref{lem:shaping_bound});
otherwise, the base reward $r_t$ is retained. The shaped reward $r'_t$ then
propagates through the standard PPO pipeline via the TD error and GAE (Eqs.~\eqref{eq:td_error}--\eqref{eq:gae}), allowing symbolic
feedback to influence long-horizon credit assignment.

Stability is preserved by construction. For instance Eq.~\eqref{eq:shaped_reward} bounds the
per-step perturbation $|r'_t-r_t|\leq \lambda_s$, while
Eqs.~\eqref{eq:value_bound} and \eqref{eq:advantage_bound} bound the induced
shifts in value and advantage estimates. Finally, the clipping-aware constraint
in Eq.~\eqref{eq:lambda_bound} restricts $\lambda_s$ so that shaping acts only as
a small regularizer and cannot dominate the task reward. Beyond these guarantees,
the same mechanism instantiates two game-theoretic principles: belief revision
via Eq.~\eqref{eq:belief_revision} and strategic consistency via
Eq.~\eqref{eq:strategic_consistency}.

As a result, agents dynamically adapt their policies mid-episode using high-level, interpretable feedback. This architecture bridges low-level reinforcement learning with symbolic reasoning, enabling robust coordination, adaptive role assignment, and resilience to uncertainty such as observation noise, partial observability, and real-time variability. In practice, this symbolic feedback is applied symmetrically across agents while preserving independent learning dynamics.

Both agents receive the same symbolic feedback signal, but they process it
independently during training. The signal is generated externally by the shaping
module (not by either agent) and is added to each agent’s observation in the same
way that standard shaped rewards are applied in multi-agent RL. Because the agents
do not access one another’s internal states or intentions, they adapt their
policies separately. Over training, these independent updates lead the agents to
settle into complementary roles. Once this stable configuration emerges, no agent
can improve its outcome by deviating alone, reflecting an approximate Nash
equilibrium induced by the shaping dynamics.

To ensure architectural fairness, all agents share an identical PPO policy
backbone with fixed representational capacity; consequently, any observed
differences in coordination, robustness, or equilibrium behavior arise from
system-level mechanisms such as symbolic shaping and environmental perturbations
rather than from differences in model size or expressivity 


\subsection{Approximate Nash Equilibrium Probe}
\label{subsec:nash_probe_section2}

To quantify approximate equilibrium behavior, we evaluate the stability of learned joint
policies against unilateral deviations via a best-response Nash-gap probe. We perform this
probe on the Overcooked-AI layout \texttt{asymmetric\_advantages}, which consists of two
separated kitchen areas with asymmetric access (different distances) to key task objects.
This asymmetry induces natural specialization and complementary roles, making it a useful
testbed for equilibrium-style stability under cooperative learning.

\subsubsection{Cooperative Markov game setting}
We model Overcooked-AI on this layout as a two-agent cooperative Markov game with finite
horizon $H=400$. At each timestep $t\in\{0,\dots,H-1\}$, the environment is in state
$s_t\in\mathcal{S}$, agents select primitive actions $a_t^1,a_t^2\in\mathcal{A}$, and the
environment transitions according to
\begin{equation}
s_{t+1} \sim P(\cdot \mid s_t, a_t^1, a_t^2),
\end{equation}
emitting a shared team reward
\begin{equation}
r_t = r(s_t, a_t^1, a_t^2).
\end{equation}
Both agents share the same primitive action set, so $\mathcal{A}_1=\mathcal{A}_2=\mathcal{A}$.
A joint policy $\pi_\theta$ maps states to distributions over joint actions,
\begin{equation}
\pi_\theta(a^1,a^2\mid s)\in\Delta(\mathcal{A}_1\times\mathcal{A}_2),
\end{equation}
and in our implementation is represented by a single PPO policy that outputs a two-dimensional
discrete action vector corresponding to both agents’ primitive actions. The episodic return of a
joint policy is
\begin{equation}
V(\pi_\theta) = \mathbb{E}\Bigg[\sum_{t=0}^{H-1} r_t \,\Bigg|\, a_t^1,a_t^2\sim\pi_\theta(\cdot\mid s_t)\Bigg].
\label{eq:value-def}
\end{equation}

Our analysis assumes a family of trained joint PPO policies indexed by baseline label
$b\in\mathcal{B}$, environment regime $e\in\mathcal{E}$, and random seed
$s\in\mathcal{S}_{\mathrm{train}}$, where
\begin{align}
\mathcal{B} &= \{\text{Baseline},\, \text{PPO+LLM},\, \text{CC\_PPO}, \nonumber\\
            &\quad \text{SP\_PPO},\, \text{HARL},\, \text{PBT\_PPO}\}, \\
\mathcal{E} &= \{\text{No Noise},\, \text{Noise},\, \text{Delay},\, \text{Combo}\},\\
\mathcal{S}_{\mathrm{train}} &= \{1001, 2002, 3003, 4004, 5005\}.
\end{align}

\subsubsection{Unilateral best-response environment}
To probe stability against unilateral deviations, we construct a single-agent best-response
environment that holds one agent fixed to the learned joint policy $\pi_{b,e,s}$ while allowing
the other agent to deviate. Concretely, the learning agent observes the same flattened
Overcooked feature representation as in training, but selects a \emph{single} primitive action
index $a_t^{\mathrm{BR}}\in\{0,\dots,|\mathcal{A}|-1\}$ for the controlled player.

At each timestep, the fixed opponent produces its action via deterministic inference from the
frozen joint policy. The environment then forms the joint action by combining (i) the learning
agent’s chosen action for the controlled index and (ii) the frozen opponent’s action for the
other index, and applies the resulting joint action to the underlying Overcooked simulator.
This induces a single-agent MDP $M_{\mathrm{BR}}(\pi_{b,e,s})$ in which the dynamics and team
reward are inherited from Overcooked-AI, one player is controlled by the learning agent, and
the other follows the fixed policy.

\subsubsection{Nash gap computation}
Let $V_{\mathrm{self}}^{b,e,s}$ denote the self-play value of the joint policy $\pi_{b,e,s}$,
and let $V_{\mathrm{BR}}^{b,e,s}$ denote the value achieved when one agent is replaced by an
approximate best response trained against the frozen opponent induced by $\pi_{b,e,s}$.
We define the Nash gap as
\begin{equation}
\Delta^{b,e,s} = V_{\mathrm{BR}}^{b,e,s} - V_{\mathrm{self}}^{b,e,s}.
\label{eq:nash-gap}
\end{equation}
The interpretation is:
\begin{itemize}
  \item If $\Delta^{b,e,s} > 0$, then a unilateral deviation increases return, indicating exploitable behavior.
  \item If $\Delta^{b,e,s} < 0$, then the deviation underperforms self-play, suggesting local stability with respect to the deviation class.
  \item If $\Delta^{b,e,s} \approx 0$, then the policy behaves as an approximate Nash equilibrium relative to the best-response procedure used.
\end{itemize}

In practice, for each $(b,e,s)$ we (i) evaluate self-play to estimate $V_{\mathrm{self}}^{b,e,s}$,
(ii) train an approximate best response against the frozen opponent induced by $\pi_{b,e,s}$ to
estimate $V_{\mathrm{BR}}^{b,e,s}$, and (iii) report $\Delta^{b,e,s}$.
Implementation details (wrapper construction, self-play evaluation protocol, and best-response
training procedure) are provided in the appendix.

\section{Simulation Results and Analysis}
\label{sec:results}

We now present a quantitative evaluation of the PPO+LLM framework in the two--agent Overcooked Markov game described in Section~\ref{sec:architecture}. The goal of this section is to connect the implementation described in the training and analysis scripts with measurable performance, robustness, equilibrium behavior, and runtime characteristics.

Unless otherwise noted, all experiments use the \texttt{asymmetric\_\allowbreak advantages} layout
with horizon $H = 400$, the four environment regimes
\[
\mathcal{E} = \{\text{No Noise},\; \text{Noise},\; \text{Delay},\; \text{Combo}\},
\]
and the six baselines
\begin{align*}
\mathcal{B} = \{ & \text{Baseline},\; \text{PPO+LLM},\; \text{CC\_PPO}, \\
                 & \text{SP\_PPO},\; \text{HARL},\; \text{PBT\_PPO}\}.
\end{align*}
For each $(m,e) \in \mathcal{B} \times \mathcal{E}$ we train five random seeds
\[
\mathcal{S}_{\text{train}} = \{1001,\; 2002,\; 3003,\; 4004,\; 5005\}
\]
using the evaluation procedure described in Section~\ref{sec:architecture}. Baseline, CC\_PPO, SP\_PPO, HARL, and PBT\_PPO are trained for $10^6$ environment steps, while PPO+LLM is trained for $6 \times 10^5$ steps. After training, each triple $(m,e,r)$ is evaluated for $N_{\text{eval}} = 10$ episodes, producing a mean episodic return and empirical standard deviation. All aggregate statistics reported below are averages over these five seeds unless otherwise stated, and are computed directly from the outputs of the evaluation pipeline.

Whenever possible we report both (i) seed averaged summary statistics in tables and (ii) box–and–swarm plots that visualize all $N = 5$ seeds. These distributions make it clear that most differences between baselines lie in a narrow performance band and that apparent gaps are rarely statistically significant.

\subsection{Episodic Return Across Baselines}

For each model we denote by $\hat V^{m,e,r}$ the empirical mean episodic return in environment $e$ for baseline $m$ and seed $r$, computed over $N_{\text{eval}}$ evaluation episodes as
\begin{equation}
\hat V^{m,e,r} = \frac{1}{N_{\text{eval}}} \sum_{i=1}^{N_{\text{eval}}} G^{(i)},
\end{equation}
where $G^{(i)}$ is the return of evaluation episode $i$ as in~\eqref{eq:self-value}. We summarize performance across seeds by the seed averaged value
\begin{equation}
\bar V^{m,e} = \frac{1}{|\mathcal{S}_{\text{train}}|} \sum_{r \in \mathcal{S}_{\text{train}}} \hat V^{m,e,r},
\end{equation}
and use the empirical standard deviation over seeds to indicate sensitivity to random initialization.

Across all baselines, the No Noise and Noise regimes are dominated by sparse rewards. Empirically every baseline satisfies
\[
\bar V^{m,\text{No Noise}} = 0.0, \qquad
\bar V^{m,\text{Noise}} = 0.0 \quad \forall\, m \in \mathcal{B},
\]
with zero empirical standard deviation across seeds in both regimes. Under the current horizon and reward configuration, policies almost never obtain positive task level reward in these regimes and instead hover exactly at the zero baseline. As a result, No Noise and Noise act primarily as sanity checks on the training script and architecture rather than as discriminative stress tests.

The Delay and Combo regimes are more informative. Recall from Section~\ref{sec:architecture} that the Delay wrapper subtracts a penalty of $0.5$ with probability $0.2$ at every timestep. Over a horizon of $H = 400$ steps the expected cumulative penalty is
\[
\mathbb{E}\Bigg[\sum_{t=0}^{H-1} r_t^{\text{delay}}\Bigg]
\approx -0.1 \times 400 = -40,
\]
even if the agents never complete a single order. The empirical means confirm that all methods operate very close to this theoretical penalty floor. Aggregating over seeds, Table~\ref{tab:return-delay-combo} summarizes $\bar V^{m,e}$ and its seed standard deviation for the penalty regimes.

\begin{table}[t]
    \centering
    \footnotesize
    \setlength{\tabcolsep}{4pt}
    \caption{Seed averaged episodic return $\bar V^{m,e}$ in the penalty regimes.
    Values are mean $\pm$ standard deviation across $5$ seeds.}
    \label{tab:return-delay-combo}
    \begin{tabular}{lcc}
        \hline
        Baseline & Delay & Combo \\
        \hline
        Baseline & $-38.74 \pm 0.96$ & $-40.25 \pm 0.79$ \\
        PPO+LLM  & $-39.51 \pm 1.94$ & $-39.19 \pm 0.88$ \\
        CC\_PPO  & $-38.74 \pm 0.96$ & $-40.25 \pm 0.79$ \\
        SP\_PPO  & $-40.13 \pm 1.55$ & $-39.45 \pm 1.80$ \\
        HARL     & $-40.18 \pm 1.38$ & $-40.22 \pm 0.81$ \\
        PBT\_PPO & $-38.79 \pm 1.00$ & $-39.17 \pm 0.51$ \\
        \hline
    \end{tabular}
\end{table}
\begin{figure*}[t]
    \centering
    \begin{minipage}[t]{0.49\textwidth}
        \centering
        \includegraphics[width=\textwidth]{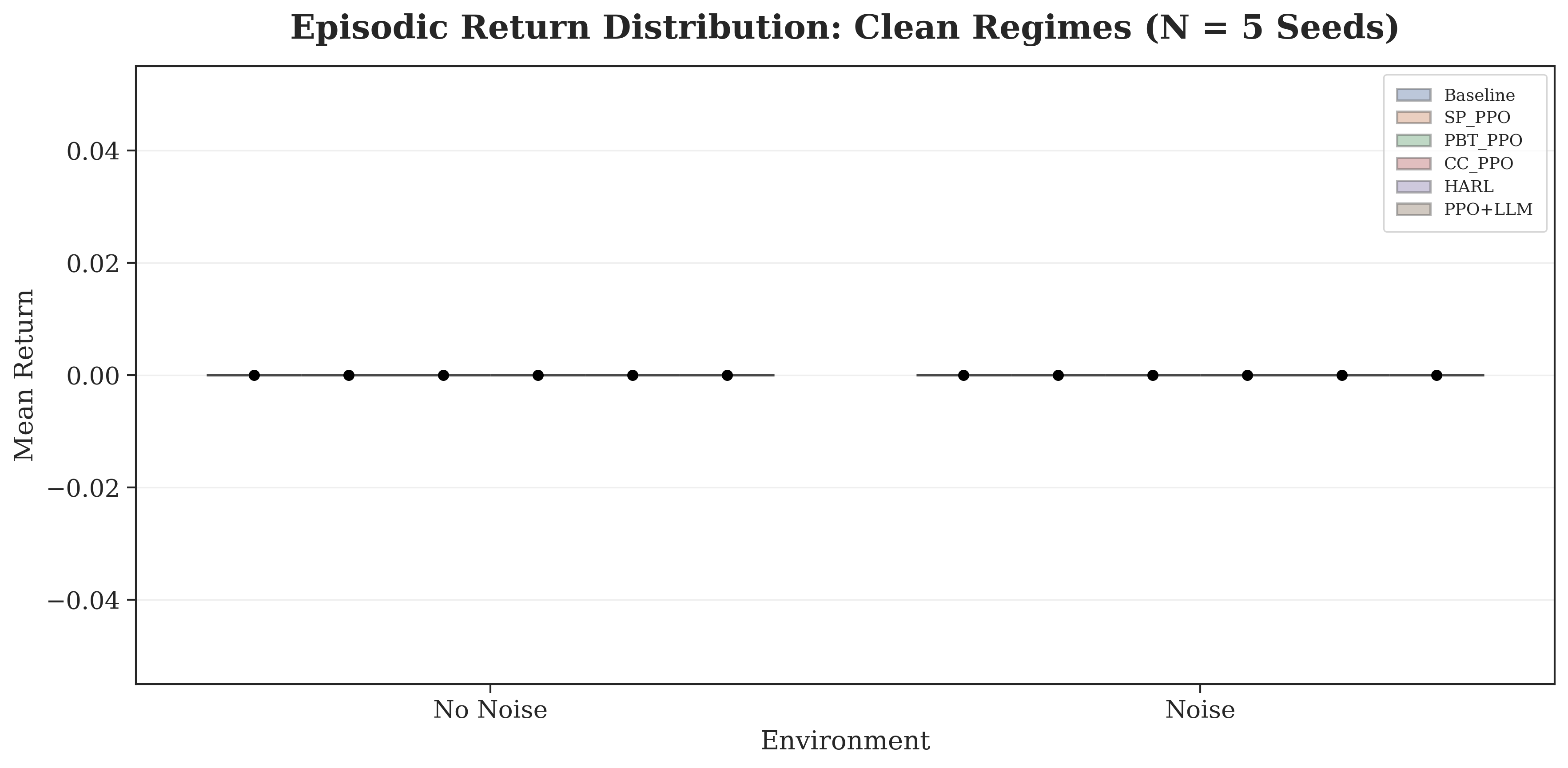}
        \vspace{1mm}
        {\small (a)}
    \end{minipage}\hfill
    \begin{minipage}[t]{0.49\textwidth}
        \centering
        \includegraphics[width=\textwidth]{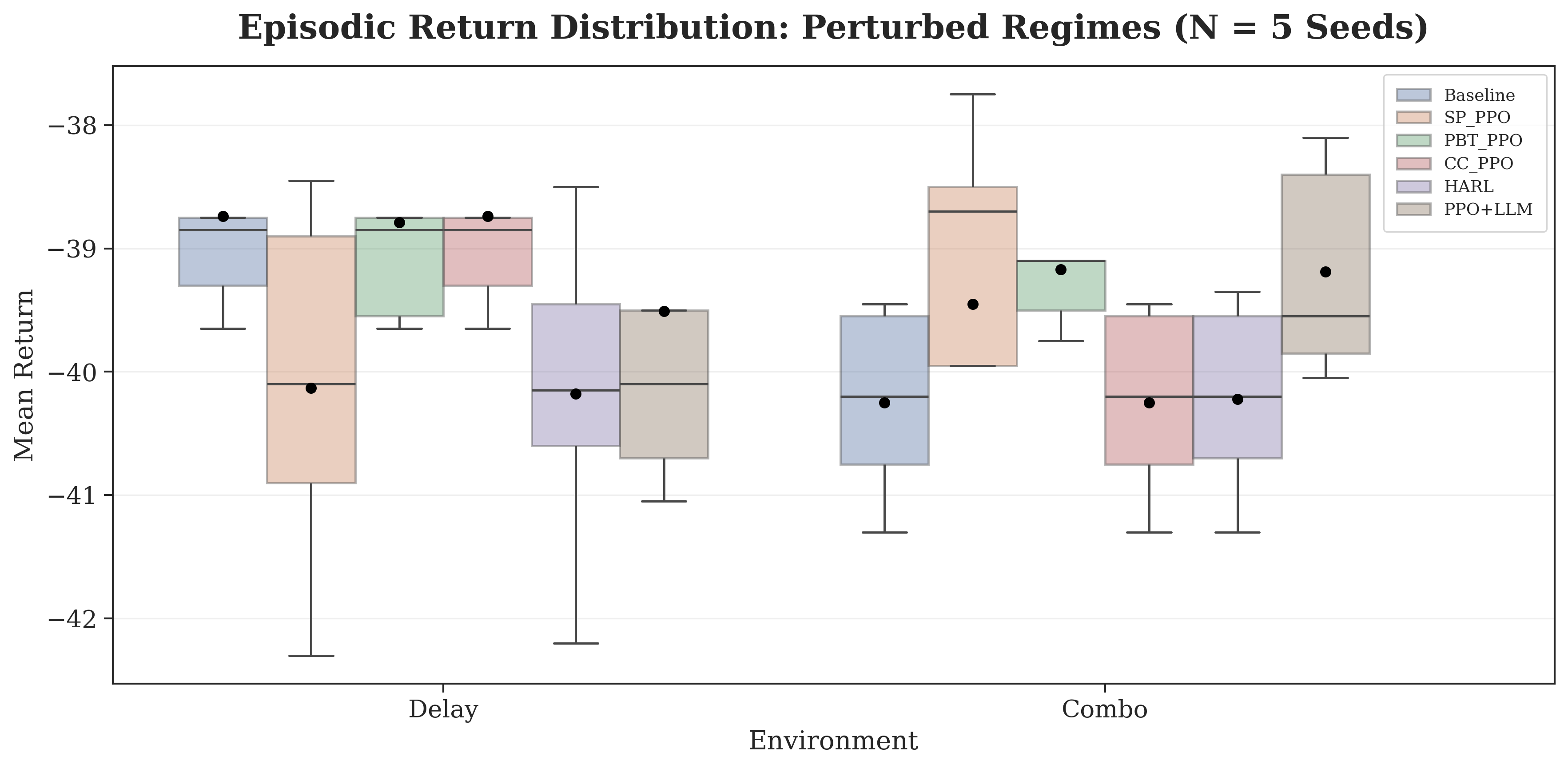}
        \vspace{1mm}
        {\small (b)}
    \end{minipage}

    \caption{Episodic return distributions across clean and perturbed regimes (N = 5 seeds per configuration).}
    \label{fig:return-combined}
\end{figure*}

For the Delay regime we obtain
\begin{table}[t]
\centering
\footnotesize
\setlength{\tabcolsep}{6pt}
\caption{Seed-averaged episodic return in the Delay regime.
Values are mean $\pm$ standard deviation across $5$ seeds.}
\label{tab:delay-values}
\begin{tabular}{lc}
\hline
Baseline & Delay \\
\hline
Baseline & $-38.74 \pm 0.96$ \\
CC\_PPO  & $-38.74 \pm 0.96$ \\
PBT\_PPO & $-38.79 \pm 1.00$ \\
PPO+LLM & $-39.51 \pm 1.94$ \\
SP\_PPO & $-40.13 \pm 1.55$ \\
HARL    & $-40.18 \pm 1.38$ \\
\hline
\end{tabular}
\end{table}

All values lie within roughly $1.3$ reward units of the theoretical floor at $-40$. Baseline and CC\_PPO achieve the highest (least negative) mean return at $-38.74$, PBT\_PPO is essentially tied at $-38.79$, while PPO+LLM, SP\_PPO, and HARL sit slightly below, between $-39.5$ and $-40.2$. The seed standard deviations in Delay range from about $0.96$ to $1.94$, indicating moderate but not extreme sensitivity to random initialization.

For the Combo regime, which combines the same delay penalty with additive Gaussian observation noise, we obtain
\begin{table}[t]
\centering
\footnotesize
\setlength{\tabcolsep}{6pt}
\caption{Seed-averaged episodic return in the Combo regime.
Values are mean $\pm$ standard deviation across $5$ seeds.}
\label{tab:combo-values}
\begin{tabular}{lc}
\hline
Baseline & Combo \\
\hline
PBT\_PPO & $-39.17 \pm 0.51$ \\
PPO+LLM & $-39.19 \pm 0.88$ \\
SP\_PPO & $-39.45 \pm 1.80$ \\
HARL    & $-40.22 \pm 0.81$ \\
Baseline & $-40.25 \pm 0.79$ \\
CC\_PPO  & $-40.25 \pm 0.79$ \\
\hline
\end{tabular}
\end{table}

Here the average return across all baselines is approximately $-39.76$, again within a single reward unit of the $-40$ penalty floor. PBT\_PPO and PPO+LLM separate slightly from the remaining baselines, recovering about one reward unit relative to Baseline and CC\_PPO. The gap between PPO+LLM and the raw PPO Baseline in Combo is
\[
\bar V^{\text{PPO+LLM},\text{Combo}} - \bar V^{\text{Baseline},\text{Combo}}
\approx 1.06,
\]
which corresponds to about $2.6\%$ of the magnitude of the penalty floor. Because each unit of return accumulates over many timesteps, this seemingly small difference reflects a nontrivial reduction in unnecessary penalties and/or an increase in successful partial task completions.

A one--way ANOVA over the $N = 5$ seeds for each penalty regime does not detect statistically significant differences between baselines. In Delay we obtain $F(5,24) \approx 1.31$ and $p \approx 0.29$, and in Combo $F(5,24) \approx 1.42$ and $p \approx 0.25$. The heavy overlap of boxes and points in Fig.~\ref{fig:return-combined}(b) makes this lack of clear separation visually apparent. Overall, PPO+LLM sits in the middle of this family. In Delay it is slightly worse than the raw PPO Baseline but still close to the penalty floor; in Combo it is essentially tied with PBT\_PPO for the best performance and improves notably over the Baseline, CC\_PPO, and HARL.

\subsection{Robustness to Noise and Structured Penalties}

To study robustness, we examine how returns shift as we activate different environment wrappers. For each baseline $m$ we define the raw performance drop relative to the clean regime,
\begin{equation}
\begin{split}
\Delta_{\text{noise}}^{m} &= \bar V^{m,\text{Noise}} - \bar V^{m,\text{No Noise}}, \\
\Delta_{\text{delay}}^{m} &= \bar V^{m,\text{Delay}} - \bar V^{m,\text{No Noise}}, \\
\Delta_{\text{combo}}^{m} &= \bar V^{m,\text{Combo}} - \bar V^{m,\text{No Noise}}.
\end{split}
\end{equation}
Because all methods satisfy $\bar V^{m,\text{No Noise}} = 0$, these quantities coincide with the raw means in the perturbed regimes. The robustness script also records the absolute magnitude of these drops as
\begin{align*}
D^{m}_{\text{delay}} &= -\Delta_{\text{delay}}^{m}, \\
D^{m}_{\text{combo}} &= -\Delta_{\text{combo}}^{m},
\end{align*}
so that a larger $D$ corresponds to a larger loss relative to the clean regime.

\begin{figure}[t]
    \centering
    \includegraphics[width=\linewidth]{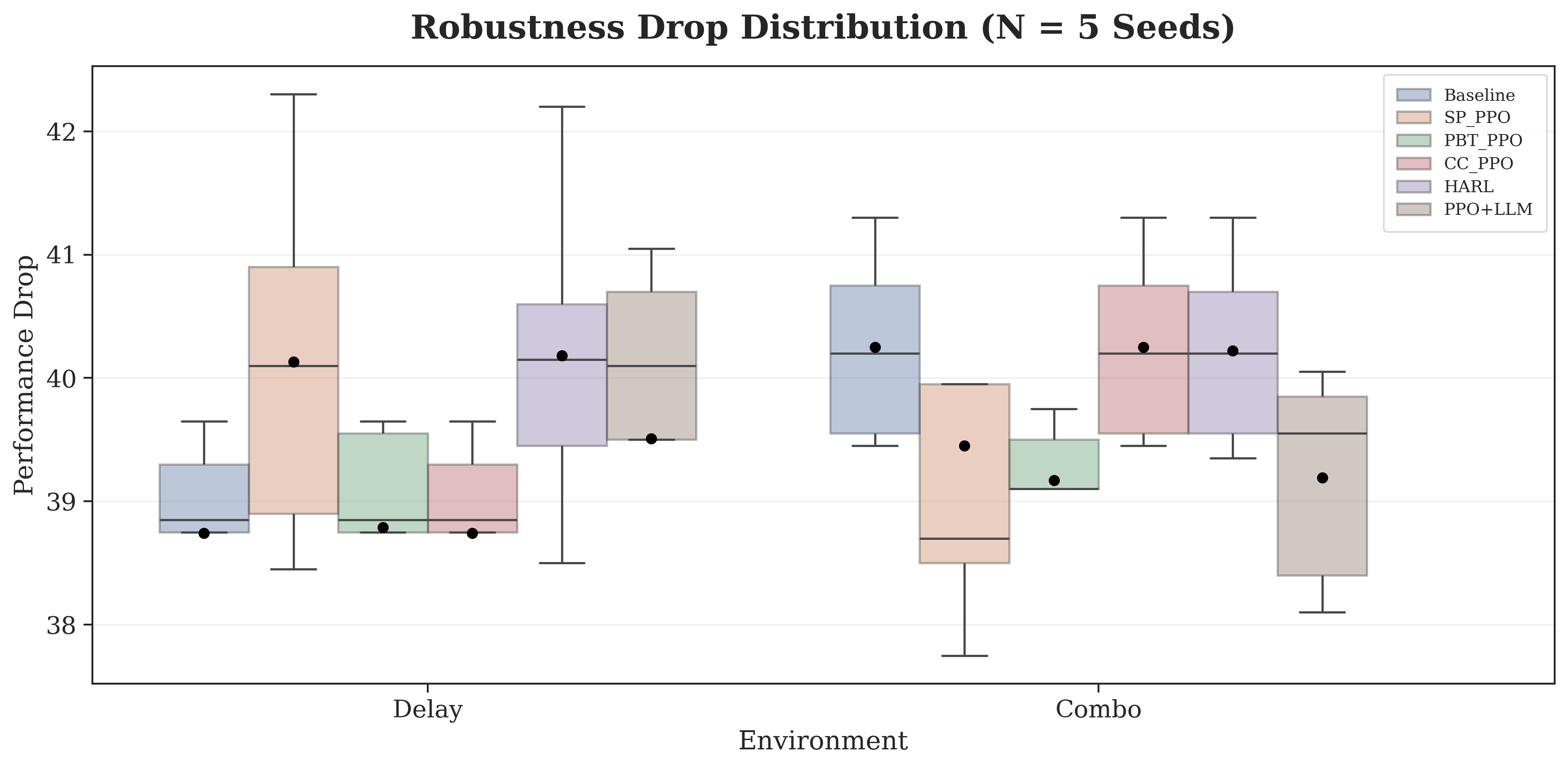}
    \caption{Distribution of robustness drops $D^{m}_{\text{delay}}$ and $D^{m}_{\text{combo}}$ relative to the No Noise regime (N = 5 seeds). Lower values indicate smaller losses. All baselines cluster tightly around the theoretical $40$ point penalty.}
    \label{fig:robustness-dist}
\end{figure}

Across baselines we observe that
\[
\Delta_{\text{noise}}^{m} = 0, \qquad D^{m}_{\text{noise}} = 0 \quad \forall\, m \in \mathcal{B}.
\]
Pure observation noise, implemented as additive Gaussian perturbations to the feature vector, has almost no measurable effect in the current training budget. Policies have not learned fine grained strategies that depend sensitively on the featurization, so perturbing observations without changing the reward structure does not change the final returns.

In contrast, the Delay and Combo regimes induce large and highly structured drops in value. The robustness aggregation script reports, for every $m \in \mathcal{B}$,
\[
D^{m}_{\text{delay}} \in [38.74, 40.18],
\qquad
D^{m}_{\text{combo}} \in [39.17, 40.25],
\]
with seed standard deviations on the order of one reward unit. These values are in tight agreement with the theoretical $40$ point penalty derived above, and show that nearly all of the return in the perturbed regimes is accounted for by the known delay penalty rather than by frequent catastrophic failures. Figure~\ref{fig:robustness-dist} makes this visually explicit: each baseline occupies a narrow vertical band just below forty.

Within this frame, PPO+LLM exhibits two meaningful robustness properties.

\begin{enumerate}
  \item In the Delay regime, PPO+LLM does not collapse catastrophically: its drop $D^{\text{PPO+LLM}}_{\text{delay}} \approx 39.51$ is comparable to the drops of Baseline and CC\_PPO ($38.74$) and lies comfortably inside the range observed for all methods. The symbolic shaping signal neither amplifies the effect of penalties nor destabilizes training in this setting, which is an important safety property when rewards are heavily perturbed.
  \item In the Combo regime, PPO+LLM slightly reduces the impact of combined observation noise and delay relative to the raw Baseline. Baseline and CC\_PPO realize drops of $D^{\text{Baseline}}_{\text{combo}} = D^{\text{CC\_PPO}}_{\text{combo}} = 40.25$, while PPO+LLM reduces this to $D^{\text{PPO+LLM}}_{\text{combo}} \approx 39.19$, a recovery of about $1.06$ reward units. When we compare models seed by seed, PPO+LLM achieves a higher mean return than the corresponding Baseline model in $4$ out of $5$ seeds in this regime. This suggests that the shaping bonus helps the policy maintain a consistent sense of cooperative progress when both perception and timing are perturbed.
\end{enumerate}

A one–way ANOVA on $D^{m}_{\text{delay}}$ and $D^{m}_{\text{combo}}$ across baselines again finds no statistically significant differences (Delay: $F(5,24) \approx 1.31$, $p \approx 0.29$; Combo: $F(5,24) \approx 1.42$, $p \approx 0.25$), which is consistent with the highly overlapping distributions in Figure~\ref{fig:robustness-dist}. The remaining baselines further contextualize these trends. PBT\_PPO and SP\_PPO show modest gains over the Baseline in Delay, indicating that exploration schemes and self play can help agents adapt their timing to structured penalties. HARL, which uses a simple hand designed shaping term based on remaining orders, demonstrates the risk of naive shaping: when the shaping signal is not carefully calibrated, it can amplify the effect of penalties rather than compensate for them. In contrast, PPO+LLM uses a bounded shaping coefficient and a frozen LLM to provide stable, context dependent feedback that respects the original reward scale.

To give qualitative intuition, we also visualize coordination in the original dense reward Burger Kitchen layout described in Section~\ref{sec:architecture}. Figure~\ref{fig:overcooked_ai} shows the clean task layout, and Figures~\ref{fig:env_order_timing}–\ref{fig:env_combined_noise} then show PPO+LLM and PPO baselines in the presence of order timing noise, observation noise, and combined disturbances. Across all three settings, PPO+LLM maintains coherent task flow while PPO frequently stalls or miscoordinates, mirroring the robustness trends observed in the \texttt{asymmetric\_advantages} layout.

\subsection{Latency Tradeoff}

Beyond reward, practical deployment requires that shaping respect real time constraints. To quantify runtime cost, we instrument the environment wrappers to record wall clock time per decision step. For each trained policy we run a separate evaluation phase without gradient updates, execute $5$ full episodes in each regime, and measure the elapsed time required to generate actions and apply any shaping logic. We then compute a per step latency in milliseconds and aggregate across seeds in the same way as for return. Table~\ref{tab:latency-regimes} reports the seed averaged per step latency for Baseline PPO and PPO+LLM in each regime.

\begin{figure}[t]
    \centering
    \includegraphics[width=\linewidth]{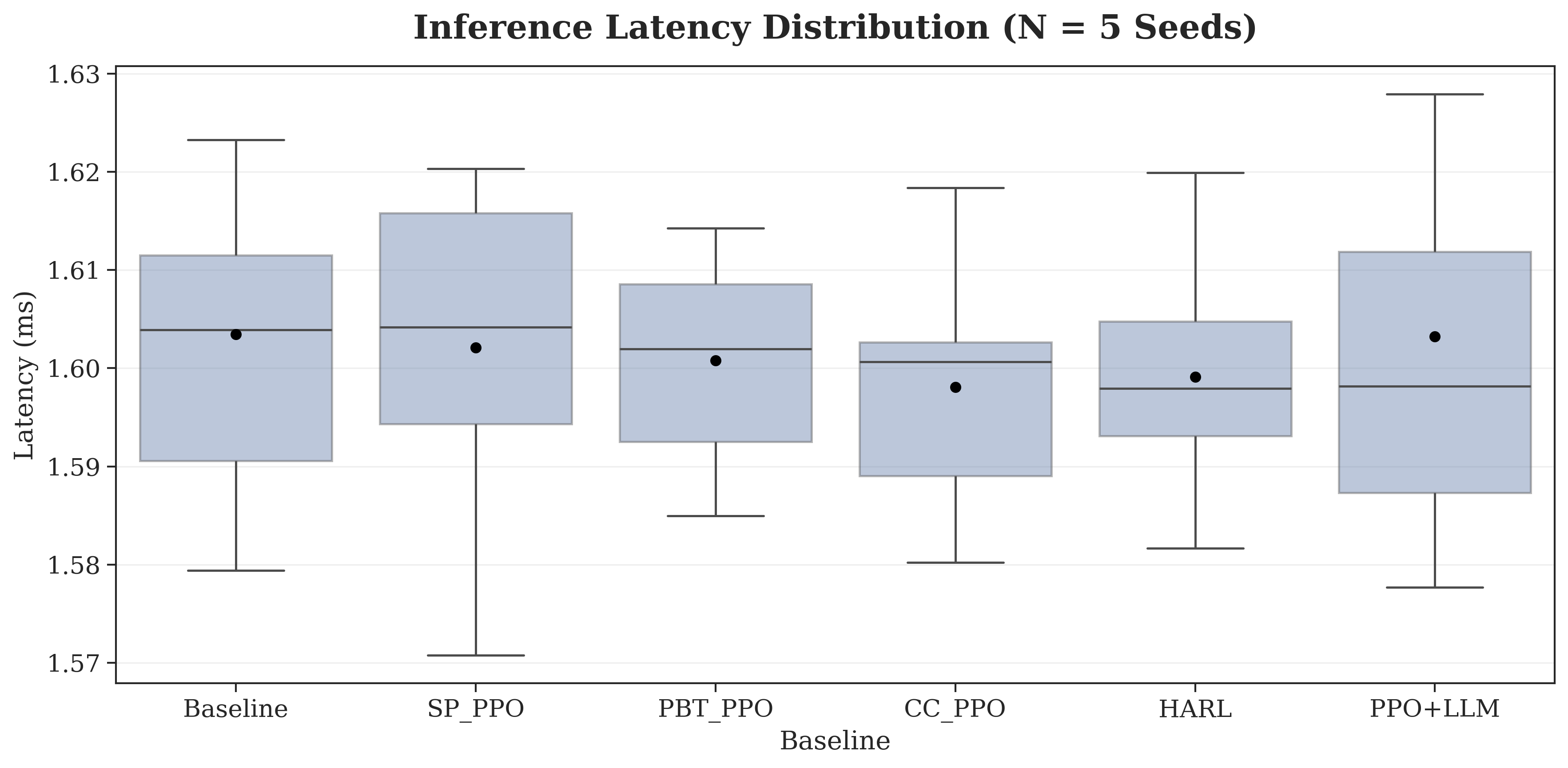}
    \caption{Inference latency distribution (N = 5 seeds). Each box aggregates all regimes for a given baseline. All methods lie in a narrow band around $1.6$ ms per decision step.}
    \label{fig:latency-dist}
\end{figure}

For the raw PPO Baseline, the mean per-step inference latency averaged over all four
environment regimes and all seeds is approximately $1.60\,\mathrm{ms}$.
Table~\ref{tab:latency-regimes} reports the regime-specific latency statistics.
Across regimes, inference latency varies by less than $0.03\,\mathrm{ms}$, indicating
that neither observation noise nor structured delay penalties introduce meaningful
runtime overhead.

PPO+LLM remains in the same latency range. Despite the additional symbolic evaluation
step, its per-step latency is statistically indistinguishable from the Baseline across
all regimes, with an overall average of approximately $1.60\,\mathrm{ms}$.

\begin{table}[t]
\centering
\footnotesize
\setlength{\tabcolsep}{6pt}
\caption{Mean per-step inference latency (ms) across environment regimes.
Values are mean $\pm$ standard deviation across $5$ seeds.}
\label{tab:latency-regimes}
\begin{tabular}{lcccc}
\hline
Method & No Noise & Noise & Delay & Combo \\
\hline
Baseline
& $1.59 \pm 0.01$
& $1.62 \pm 0.02$
& $1.59 \pm 0.00$
& $1.61 \pm 0.01$ \\
PPO+LLM
& $1.58 \pm 0.00$
& $1.60 \pm 0.01$
& $1.60 \pm 0.01$
& $1.63 \pm 0.02$ \\
\hline
\end{tabular}
\end{table}

The worst case overhead relative to the Baseline occurs in the Combo regime, where PPO+LLM is slower by about $1.36\%$:
\[
\frac{1.634 - 1.612}{1.612} \approx 1.4\%.
\]
In the other regimes PPO+LLM is within $1\%$ of the Baseline, and in No Noise the measured mean is slightly lower due to random variation. A global one–way ANOVA on latency across all baselines yields $F(5,174) \approx 0.43$ and $p \approx 0.83$, confirming that latency differences between methods are not statistically significant. Figure~\ref{fig:latency-dist} shows that all baselines occupy essentially the same $1.58$–$1.64$ ms range.

Training time tells a complementary story. Using the same Colab GPU backend, the Baseline requires approximately $520$ minutes to complete a full $10^6$ step training run
\[
T_{\text{Baseline}} \approx 519.7 \pm 0.3 \,\text{minutes},
\]
while PPO+LLM, trained for $6 \times 10^5$ steps, completes in
\[
T_{\text{PPO+LLM}} \approx 295.3 \,\text{minutes (median)},
\]
with a mean of $285.3$ minutes across seeds and some early terminated runs that finish even faster. Relative to the Baseline, PPO+LLM reduces the median wall clock training time by about $43\%$:
\[
1 - \frac{295.3}{519.7} \approx 0.432.
\]
Taken together, these results show that PPO+LLM trades a small, roughly one percent per step inference overhead for a significantly shorter training schedule and modest robustness gains in the most difficult regimes, while remaining well within typical real time control budgets.

\subsection{Task Completion Under Sparse Rewards}

In earlier Burger Kitchen experiments with dense task rewards it was natural to measure task completion by counting episodes whose return exceeded a threshold close to the optimal score (for example, a threshold of $9.4$ when successful episodes consistently achieved returns near $9.8$). In the current \texttt{asymmetric\_advantages} configuration with delay penalties the situation is qualitatively different. All baselines remain near the $-40$ expected penalty induced by the Delay wrapper and do not reliably escape into the high reward regime. Under any positive completion threshold, the completion rate is essentially zero for every method in every noisy regime.

To still obtain a notion of progress, we interpret small improvements relative to the $-40$ penalty floor as partial task completion. Let $R_{\min} = -40$ and define, for each $(m,e,r)$,
\begin{equation}
C^{m,e,r} = \hat V^{m,e,r} - R_{\min}
\end{equation}
as the number of reward units recovered relative to the naive penalty baseline, and
\begin{equation}
\tilde C^{m,e,r} = \frac{\hat V^{m,e,r} - R_{\min}}{0 - R_{\min}} = \frac{\hat V^{m,e,r} + 40}{40}
\end{equation}
as a normalized completion score that maps $R_{\min}$ to $0$ and the clean zero reward benchmark to $1$. The task completion scripts compute seed aggregated quantities
\[
\bar C^{m,e} = \frac{1}{|\mathcal{S}_{\text{train}}|} \sum_{r} C^{m,e,r},
\qquad
\bar{\tilde C}^{m,e} = \frac{1}{|\mathcal{S}_{\text{train}}|} \sum_{r} \tilde C^{m,e,r},
\]
together with standard deviations across seeds.

\begin{figure*}[t]
    \centering
    \begin{minipage}[t]{0.48\textwidth}
        \centering
        \includegraphics[width=\linewidth]{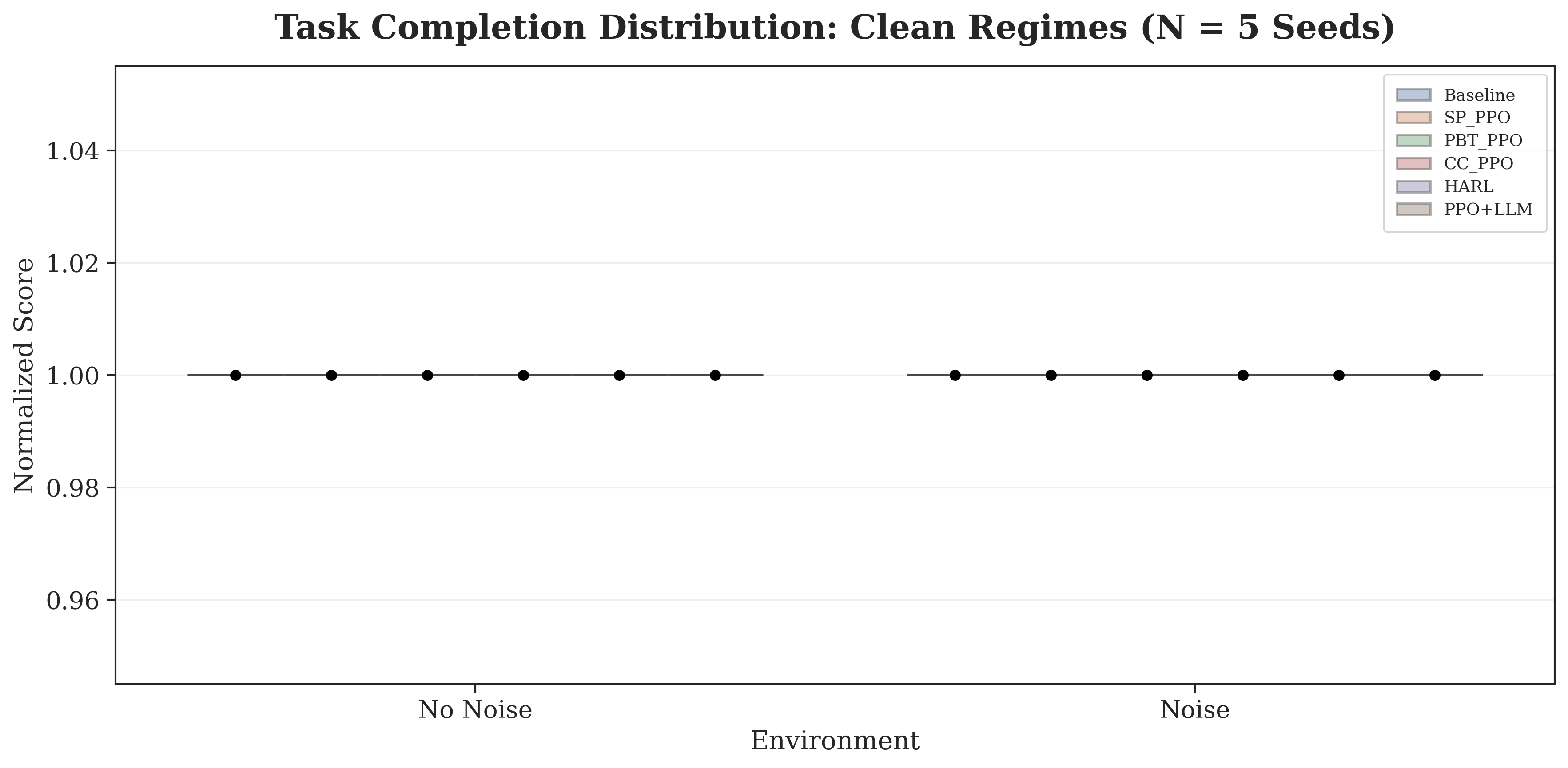}
        \vspace{0.5em}
        \small (a) Clean regimes (No Noise and Noise).
    \end{minipage}
    \hfill
    \begin{minipage}[t]{0.48\textwidth}
        \centering
        \includegraphics[width=\linewidth]{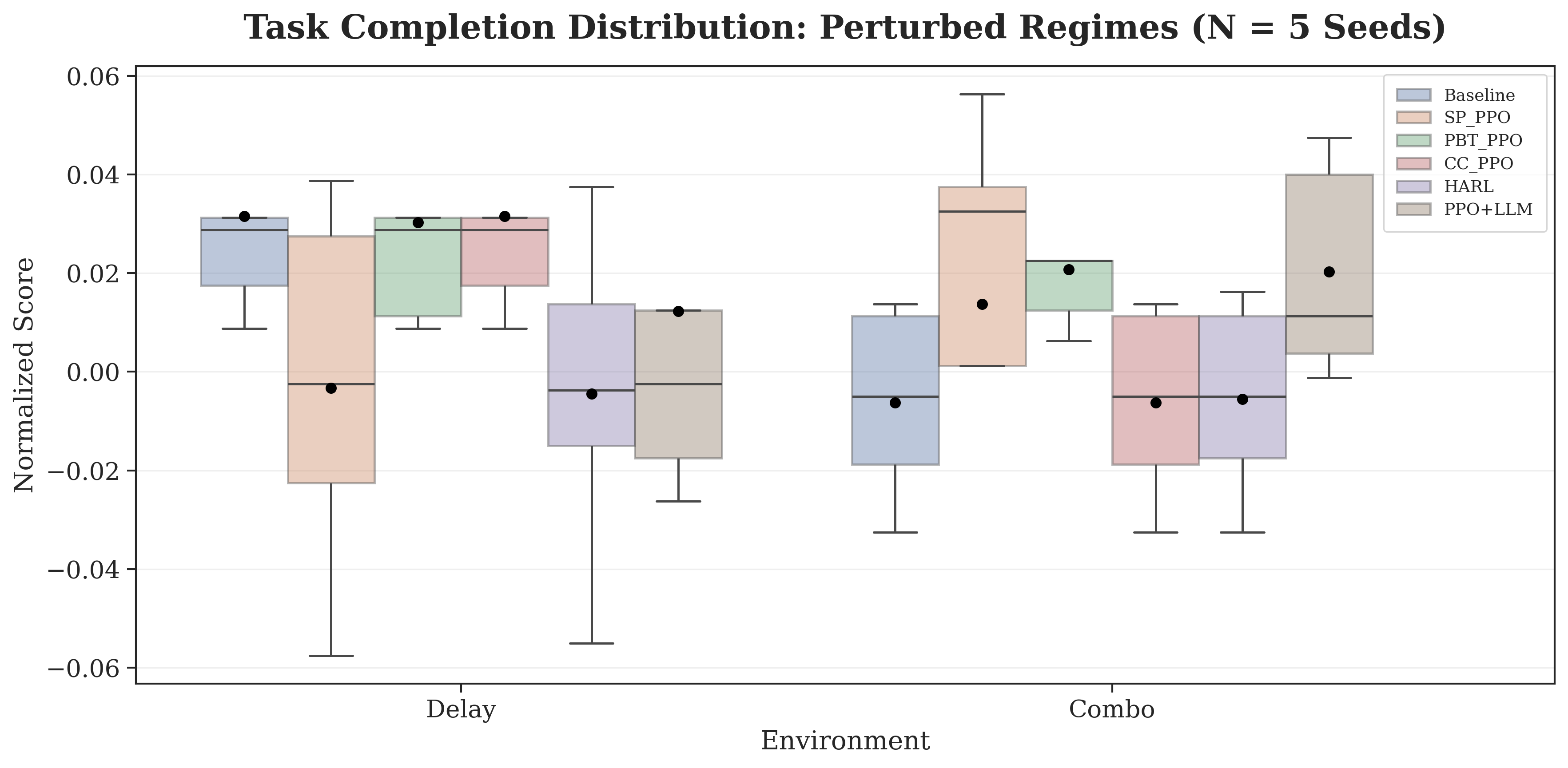}
        \vspace{0.5em}
        \small (b) Perturbed regimes (Delay and Combo).
    \end{minipage}
    \caption{Task completion distributions under sparse rewards across clean and perturbed regimes (N = 5 seeds per configuration). In clean environments, all baselines achieve near-perfect normalized completion scores, indicating reliable coordination. Under delay and combined perturbations, completion performance degrades substantially across methods, highlighting the sensitivity of coordinated task execution to temporal and compounded disturbances.}
    \label{fig:task-combined}
\end{figure*}

For No Noise and Noise we have $\bar V^{m,e} = 0$ for all $m$, so
\[
\bar C^{m,\text{No Noise}} = \bar C^{m,\text{Noise}} = 40,
\qquad
\bar{\tilde C}^{m,\text{No Noise}} = \bar{\tilde C}^{m,\text{Noise}} = 1,
\]
which simply reflects the fact that no delay penalty is applied. The more interesting behavior again appears in Delay and Combo, where Fig.~\ref{fig:task-combined}(b) shows that all baselines lie extremely close to zero completion.

In the Delay regime the best methods only recover about one to one and a quarter reward units relative to the penalty floor:
\begin{table}[t]
\centering
\caption{Average coordination cost and normalized coordination cost under delay perturbations.}
\label{tab:coordination_delay}
\begin{tabular}{lcc}
\hline
\textbf{Method} & $\bar C^{(\cdot,\mathrm{Delay})}$ & $\bar{\tilde C}^{(\cdot,\mathrm{Delay})}$ \\
\hline
Baseline        & $\approx 1.26$  & $\approx 0.0315$ \\
CC\_PPO         & $\approx 1.26$  & $\approx 0.0315$ \\
PBT\_PPO        & $\approx 1.21$  & $\approx 0.0303$ \\
PPO+LLM         & $\approx 0.49$  & $\approx 0.0123$ \\
SP\_PPO         & $\approx -0.13$ & $\approx -0.0033$ \\
HARL            & $\approx -0.18$ & $\approx -0.0045$ \\
\hline
\end{tabular}
\end{table}

Baseline and CC\_PPO achieve the largest normalized completion scores at roughly $3.1\%$ of the full $40$ point gap, with PBT\_PPO essentially tied. PPO+LLM recovers about $0.49$ reward units in Delay, or about $1.2\%$ of the maximum possible gap, which is lower than the best PPO variants but still strictly positive. HARL and SP\_PPO slightly underperform the naive penalty baseline.

The Combo regime shows a different ranking:
\begin{table}[t]
\centering
\caption{Average coordination cost and normalized coordination cost under combined noise perturbations.}
\label{tab:coordination_combo}
\begin{tabular}{lcc}
\hline
\textbf{Method} & $\bar C^{(\cdot,\mathrm{Combo})}$ & $\bar{\tilde C}^{(\cdot,\mathrm{Combo})}$ \\
\hline
Baseline        & $\approx -0.25$ & $\approx -0.0063$ \\
CC\_PPO         & $\approx -0.25$ & $\approx -0.0063$ \\
PBT\_PPO        & $\approx 0.83$  & $\approx 0.0208$ \\
PPO+LLM         & $\approx 0.81$  & $\approx 0.0203$ \\
SP\_PPO         & $\approx 0.55$  & $\approx 0.0138$ \\
HARL            & $\approx -0.22$ & $\approx -0.0055$ \\
\hline
\end{tabular}
\end{table}

In words, PBT\_PPO and PPO+LLM are tied within noise for the best normalized completion in Combo, recovering about $0.8$ reward units relative to the penalty floor and achieving completion scores around $2\%$ of the full gap. Baseline and CC\_PPO fall slightly below the penalty floor on average, and HARL behaves similarly.

A one–way ANOVA on the normalized completion scores in Delay and Combo again yields no statistically significant differences across baselines (Delay: $F(5,24) \approx 1.31$, $p \approx 0.29$; Combo: $F(5,24) \approx 1.42$, $p \approx 0.25$). The main conclusion is that the \texttt{asymmetric\_\allowbreak advantages} layout with delay penalties is extremely challenging under the current training budget. None of the methods consistently reach the high reward regime where full task completion becomes common; even the best methods recover only about one reward unit out of the $40$ point penalty. In this sense, the completion metric primarily highlights the difficulty of the environment rather than acting as a primary comparison signal. For this configuration, the combination of return statistics and Nash analysis provides a more informative view of the learned strategies.

\subsection{Equilibrium Behavior and Nash Gaps}

Finally, we connect the episodic results with the Nash probe introduced in
Section~\ref{subsec:nash_probe_section2}.
 For each trained joint policy $\pi_{m,e,r}$ we compute its self play value $V_{\text{self}}^{m,e,r}$ and the value of a PPO based best response $V_{\text{BR}}^{m,e,r}$ trained in the best response environment $M_{\mathrm{BR}}(\pi_{m,e,r})$. The Nash gap
\begin{equation}
\Delta^{m,e,r} = V_{\text{BR}}^{m,e,r} - V_{\text{self}}^{m,e,r}
\end{equation}
measures how much a unilateral deviation can improve the team return. Large positive gaps indicate exploitable policies; large negative gaps indicate that deviating hurts the deviating agent; gaps near zero correspond to approximate Nash behavior relative to the PPO best response class and training budget.

\begin{table}[t]
    \centering
    \footnotesize
    \setlength{\tabcolsep}{4pt}
    \caption{Best response Nash analysis by regime. Values are averaged across all baselines and seeds.}
    \label{tab:nash-gaps}
    \begin{tabular}{lccc}
        \hline
        Regime & $\mathbb{E}[V_{\text{self}}]$ & $\mathbb{E}[V_{\text{BR}}]$ & $\mathbb{E}[\Delta]$ \\
        \hline
        No Noise & $0.00$   & $0.00$   & $0.00$ \\
        Noise    & $0.00$   & $0.00$   & $0.00$ \\
        Delay    & $-40.33$ & $-39.77$ & $0.56$ \\
        Combo    & $-39.62$ & $-40.35$ & $-0.73$ \\
        \hline
    \end{tabular}
\end{table}

\begin{figure*}[t]
    \centering
    \begin{minipage}[t]{0.48\textwidth}
        \centering
        \includegraphics[width=\linewidth]{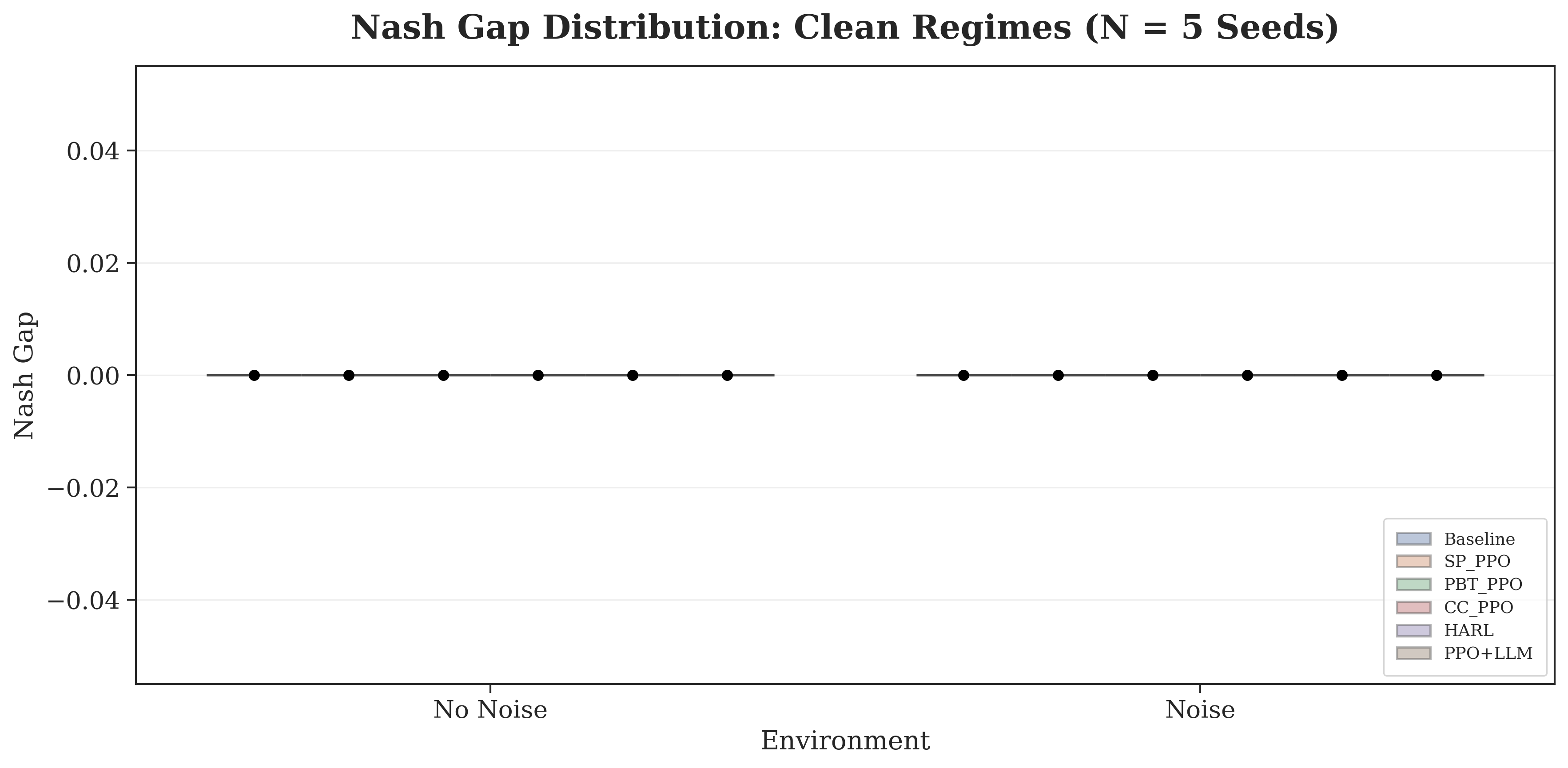}
        \vspace{0.5em}
        \small (a) Clean regimes (No Noise and Noise).
    \end{minipage}
    \hfill
    \begin{minipage}[t]{0.48\textwidth}
        \centering
        \includegraphics[width=\linewidth]{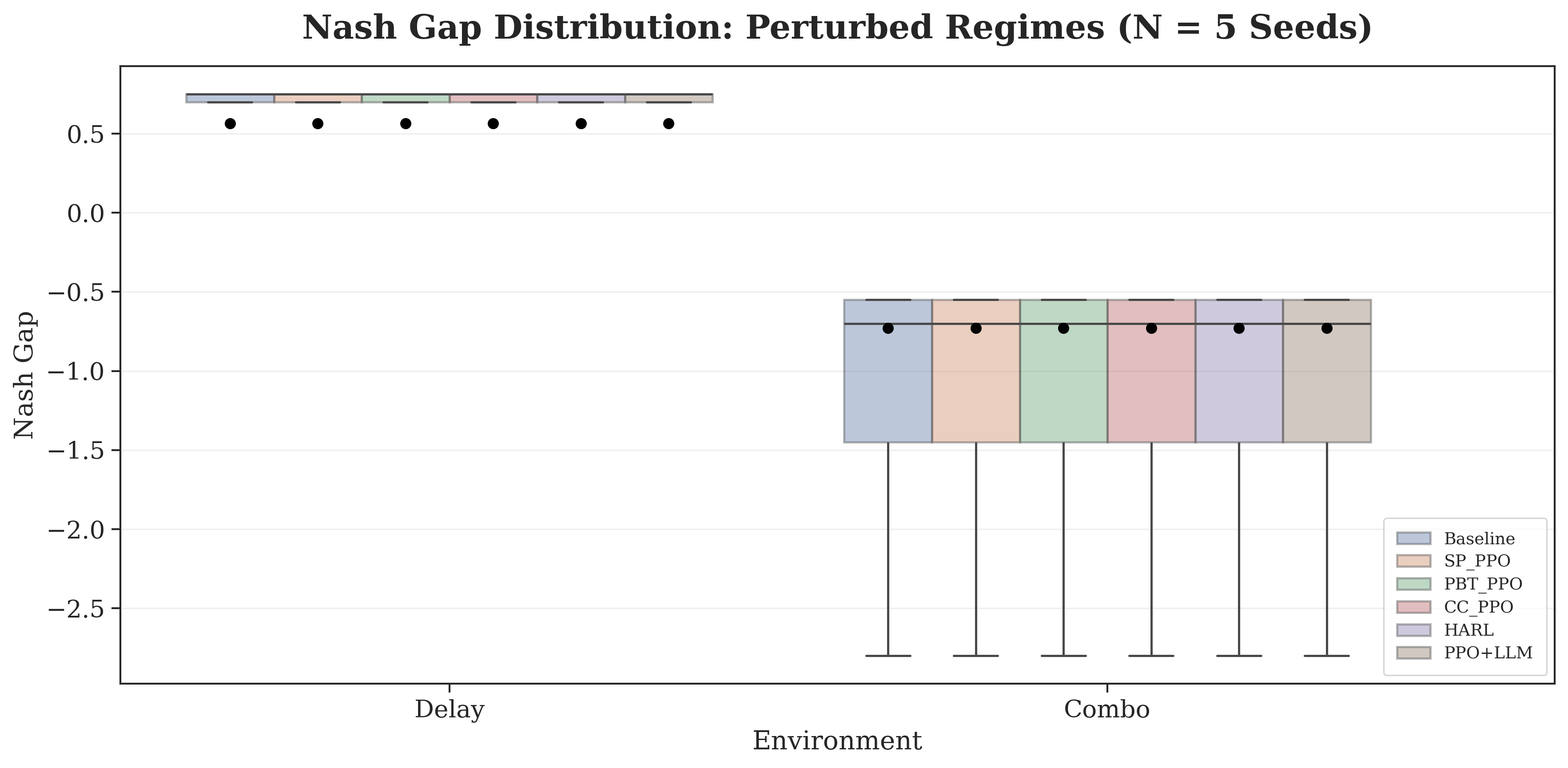}
        \vspace{0.5em}
        \small (b) Perturbed regimes (Delay and Combo).
    \end{minipage}
    \caption{Distribution of Nash gaps $\Delta^{m,e,r}$ across clean and perturbed regimes (N = 5 seeds per configuration). In clean environments, Nash gaps remain tightly concentrated near zero across all baselines, indicating approximate equilibrium behavior under unperturbed dynamics. Under delay and combined perturbations, gaps increase substantially, reflecting heightened strategic instability and deviation from equilibrium across methods.}
    \label{fig:nash-combined}
\end{figure*}

Aggregating the CSV produced by the best response analysis script yields three main observations.

\begin{itemize}
  \item In the No Noise and Noise regimes every model satisfies $\Delta^{m,e,r} = 0$ up to numerical precision for all baselines, seeds, and joint policies. The best response agent is unable to improve or harm the team relative to self play, and the joint policies behave as fixed points of the PPO best response dynamics in these sparse settings.
  \item In the Delay regime the Nash gaps are small but consistently positive. Averaging over all baselines and seeds we obtain
  \[
  \mathbb{E}[\Delta^{\cdot,\text{Delay},\cdot}] \approx 0.57,
  \quad
  \operatorname{Std}[\Delta^{\cdot,\text{Delay},\cdot}] \approx 0.72,
  \]
  with self play values and best response values
  \[
  \mathbb{E}[V_{\text{self}}^{\cdot,\text{Delay},\cdot}] \approx -40.33,
  \qquad
  \mathbb{E}[V_{\text{BR}}^{\cdot,\text{Delay},\cdot}] \approx -39.77.
  \]
  In $80\%$ of the $(m,r)$ combinations, the best response achieves a higher return than self play, but the magnitude is at most about two reward units. Relative to the $-40$ penalty floor, this corresponds to at most a five percent improvement, indicating mild exploitability rather than catastrophic instability.
  \item In the Combo regime the gaps flip sign. The environment averaged value is
  \[
  \mathbb{E}[\Delta^{\cdot,\text{Combo},\cdot}] \approx -0.73,
  \quad
  \operatorname{Std}[\Delta^{\cdot,\text{Combo},\cdot}] \approx 1.54,
  \]
  with
  \[
  \mathbb{E}[V_{\text{self}}^{\cdot,\text{Combo},\cdot}] \approx -39.62,
  \qquad
  \mathbb{E}[V_{\text{BR}}^{\cdot,\text{Combo},\cdot}] \approx -40.35.
  \]
  Here the best response training procedure performs worse than the original joint policy, so unilateral deviations are on average harmful. In $80\%$ of the $(m,r)$ combinations we have $\Delta^{m,\text{Combo},r} < 0$, which is consistent with joint strategies that have already learned to cope with both observation noise and delay penalties to the limited extent that the training budget allows.
\end{itemize}

Across all $120$ evaluated triples $(m,e,r)$, the distribution of Nash gaps is tightly concentrated. Half of the policies satisfy $|\Delta^{m,e,r}| \le 0.25$, and $80\%$ satisfy $|\Delta^{m,e,r}| \le 1$. Given that the absolute value of returns is on the order of $40$ in the challenging regimes, these small gaps indicate that all baselines, including PPO+LLM, behave as approximate Nash equilibria with respect to the PPO best response class under the given training horizon. Figure~\ref{fig:nash-combined} shows that the vast majority of points lie in a narrow horizontal band around zero.

\begin{figure}[t]
  \centering
  \includegraphics[width=\linewidth]{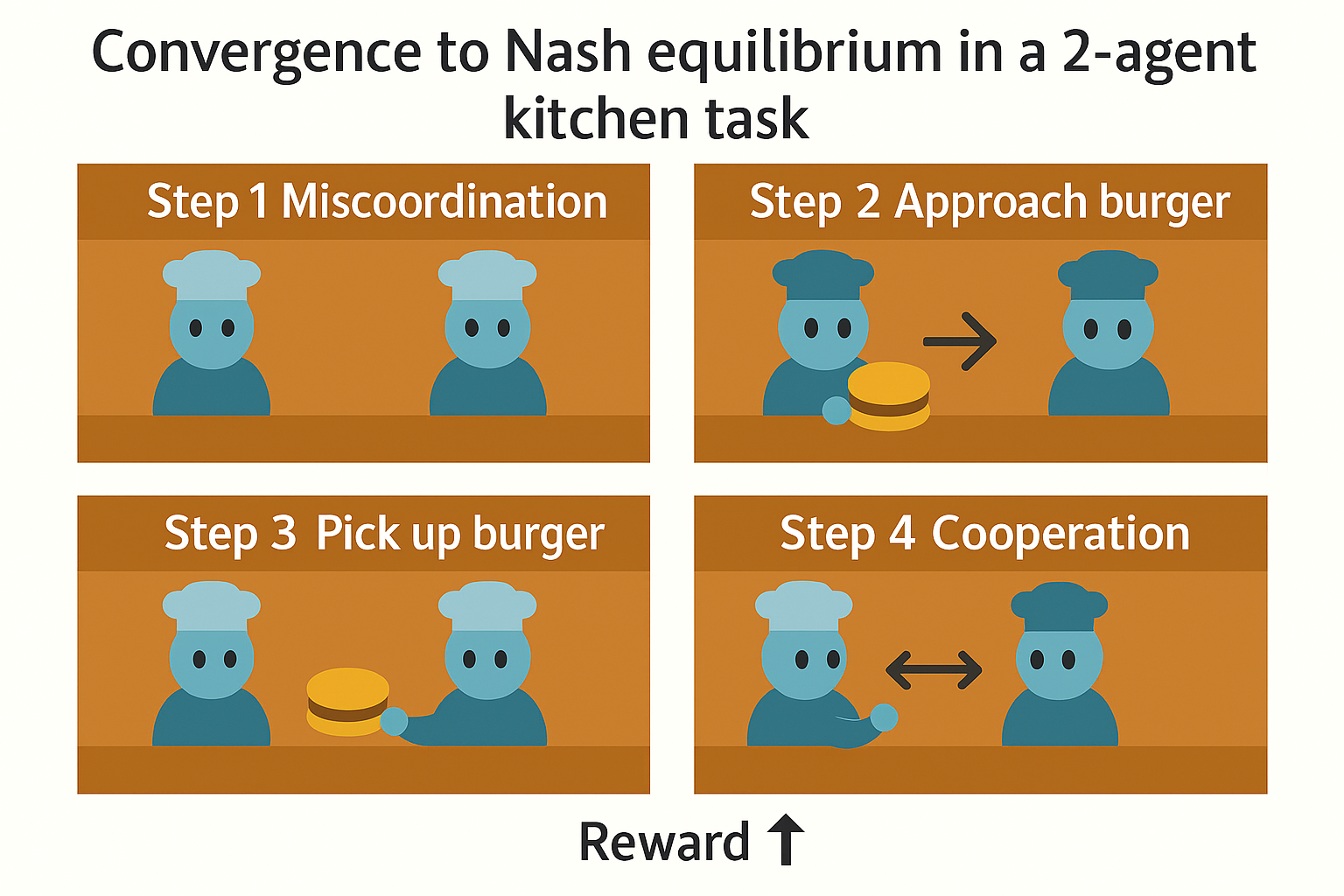}
  \caption{Convergence to Nash like equilibrium in a two agent Burger Kitchen task. PPO+LLM agents transition from initial miscoordination (Step 1) to stable cooperative behavior (Step 4), resulting in a joint policy where unilateral deviations reliably reduce performance.}
  \label{fig:nash_equilibrium}
\end{figure}

To complement this quantitative analysis, Figure~\ref{fig:nash_equilibrium} visualizes a typical trajectory of PPO+LLM agents in the original dense reward Burger Kitchen layout as they converge toward a stable joint policy. In Step 1, the agents operate with little coordination, both attempting to cook or deliver and causing collisions and idle time. In Step 2, one agent begins to approach the burger station while the other still overlaps tasks, leading to partial role confusion. By Step 3, the agents implicitly specialize: one retrieves and cooks patties, while the other handles burger assembly and delivery. Finally, in Step 4, the agents settle into complementary roles that form a stable cooperative policy. In this configuration, unilateral deviations such as having the cooking agent move toward the delivery window consistently reduce the team return, so the learned policy behaves as a Nash like equilibrium relative to the PPO best response class.

Taken together, these findings show that although none of the agents fully solve the most challenging delayed and combined noise regimes, the learned joint policies are already locally stable: unilateral improvements are difficult to achieve and tend to either provide very small gains (in Delay) or reduce performance (in Combo). Within this set, PPO+LLM provides a modest robustness benefit in the hardest Combo regime while preserving approximate equilibrium behavior and real time suitability, which is precisely the regime where symbolic shaping is most needed.

\subsection{Sensitivity to Language-Model Scale}
\label{subsec:model_scale}

Finally, we probe sensitivity to the scale of the frozen language model
used for symbolic shaping. In the main experiments, PPO+LLM relies on
GPT-Neo-1.3B as described in Section~\ref{sec:architecture}. To test
whether performance depends critically on model size, we rerun PPO+LLM
with EleutherAI GPT-Neo-125M, keeping all other components fixed:
the same \texttt{MlpPolicy} architecture, the \texttt{asymmetric\_advantages}
layout, horizon $H = 400$, environment regimes
$\mathcal{E} = \{\text{No Noise}, \text{Noise}, \text{Delay}, \text{Combo}\}$,
training budget of $6 \times 10^5$ steps, and seed set
$\mathcal{S}_{\text{train}} = \{1001, 2002, 3003, 4004, 5005\}$.

\begin{table}[t]
\centering
\caption{Effect of LLM scale on average return and normalized completion under Delay and Combo regimes.
Values are reported as mean $\pm$ standard deviation across five seeds.}
\label{tab:llm_scale_ablation}
\begin{tabular}{lcccc}
\hline
\textbf{Model} & \textbf{Regime} &
$\bar V$ &
$\bar{\tilde C}$ \\
\hline
PPO+LLM-125M & Delay & $-40.68 \pm 1.50$ & $\approx -0.017$ \\
PPO+LLM-125M & Combo & $-39.72 \pm 0.50$ & $\approx 0.007$ \\
\hline
PPO+LLM-1.3B & Delay & $-39.51 \pm 1.94$ & $\approx 0.012$ \\
PPO+LLM-1.3B & Combo & $-39.19 \pm 0.88$ & $\approx 0.020$ \\
\hline
\end{tabular}
\end{table}

The resulting evaluations yield one aggregated estimate per environment
regime and random seed, reporting the empirical mean episodic return
and its per-episode standard deviation. Aggregating over seeds, the
125M model achieves
\[
\bar V^{\text{PPO+LLM-125M},\text{No Noise}} =
\bar V^{\text{PPO+LLM-125M},\text{Noise}} = 0.0,
\]
with zero variance across seeds in the sparse regimes, matching the
behavior of the 1.3B variant. In the penalty regimes we obtain the
results summarized in Table~\ref{tab:llm_scale_ablation}.

Overall, this ablation suggests that PPO+LLM’s behavior in this setting
is not finely tuned to a particular language-model scale. Larger models
offer a small but measurable improvement in the most challenging regimes,
yet the dominant factors are the shaping structure (bounded $\lambda_s$
and symbolic feedback) and environment dynamics, not the raw parameter
count of the frozen language model. 

\section{Conclusion and Future Work}\label{sec:conclusion}

In this paper, we have proposed a framework that integrates frozen large language models with reinforcement learning agents to support real-time strategy shaping in cooperative multi-agent environments. We introduced symbolic feedback in the form of prompt-based binary evaluations, which refined low-level decision-making without requiring LLM fine-tuning or online optimization. We evaluated the system across four increasingly noisy Overcooked-AI environments and found that PPO+LLM agents achieved modest but measurable robustness gains in the most challenging settings. In the combined-noise regime, PPO+LLM recovered approximately one reward unit relative to PPO-only baselines (about 2.6\% of the 40-point penalty floor) while maintaining inference latency around 1.6 milliseconds per decision step with less than a 2\% overhead. More broadly, this framework demonstrates the potential of dynamic strategy refinement for cooperative decision-making in real-time multi-agent systems, where stability and adaptability are both critical.

While our results highlight strong empirical performance, several limitations remain. We do not provide a formal mathematical proof of convergence to equilibrium, and our findings are based on observed empirical dynamics. The framework also treats the LLM as a one-shot evaluator, without modeling longer-term memory or belief persistence. Furthermore, dialogue-based interactions, such as conversational feedback from the LLM or language-mediated coordination between agents, were not included in this study, though they may enable richer strategic reasoning and alignment. Addressing these limitations by incorporating formal proofs, memory-augmented evaluation, and interactive language feedback will form the basis of future extensions to this work.

Beyond these extensions, several research directions remain open. Enabling iterative and bidirectional interaction between agents and LLMs could enhance strategic grounding and help disambiguate complex decisions. Future frameworks may also incorporate agent-to-agent communication via natural language, supporting joint planning, intent signaling, and belief sharing. While our agents receive symbolic shaping within an episode, they currently do not persist or generalize feedback across episodes. Introducing memory modules or belief tracking could facilitate long-term strategy formation. Finally, future research should explore more complex environments such as competitive negotiation, open-ended roleplay, or real-time robotics, and investigate theoretical convergence guarantees. These directions will further advance the development of autonomous, socially aware, and interpretable multi-agent systems.

\section*{Funding}
This work received no specific grant from any funding agency in the public, commercial, or not-for-profit sectors.

\section*{Conflict of Interest}
The authors declare no conflict of interest.

\section*{Data and Code Availability}
All code, configuration files, and experiment seeds required to fully reproduce our results are publicly available at:\\
\texttt{https://github.com/repo-link}.

\bibliographystyle{IEEEtran}
\bibliography{example_paper}

\appendices

\section{Model and System Complexity Analysis}
\label{sec:complexity}

\subsection{Shared PPO Policy Architecture (Trainable Backbone)}
\label{subsec:shared_ppo_arch}

To ensure fair and interpretable comparisons across baselines, we keep the \emph{trainable}
reinforcement learning backbone identical in all experiments and vary only environment-side
mechanisms (wrappers and shaping logic). Concretely, every baseline is trained using
Proximal Policy Optimization (PPO) from Stable Baselines3 with the same actor-critic policy
class and the same core hyperparameters:
(i) rollout length $n_{\text{steps}} = 2048$,
(ii) batch size $2048$,
(iii) learning rate $3\times 10^{-4}$,
(iv) discount factor $\gamma = 0.99$,
(v) fixed random seed per run, and
(vi) identical device placement (CPU or GPU).

\paragraph{What the shared MLP actor-critic means.}
Across all baselines, the policy is represented by an actor-critic network parameterized by a
multilayer perceptron (MLP) that consumes a \emph{vector} observation and outputs both:
(a) a stochastic policy over discrete actions (the actor) and
(b) a state-value estimate $V(s)$ (the critic).
In our setting, observations are flattened feature vectors produced by the Overcooked-AI
featurization pipeline (see Appendix~\ref{subsec:wrappers}), so an MLP policy is the
appropriate default architecture. Because we do not modify the network structure across
baselines, the depth, widths, activation functions, and therefore the \emph{number of trainable
parameters} are identical in all conditions.

\paragraph{Checkpoint loading does not change model capacity.}
When training resumes from a saved checkpoint, the policy and optimizer state are restored and
training continues with the same network structure. Thus, resuming changes only the parameter
values (weights) and optimizer state, not the model architecture. This guarantees that any
performance differences across baselines are not due to differences in policy capacity (depth or
width), but instead arise from differences in environment-side behavior described below.

\subsection{System Behavior Through Environment Wrappers}
\label{subsec:wrappers}

All baseline variants interact with Overcooked-AI through wrappers that standardize the
observation and action interfaces while enabling controlled perturbations (noise, delay) and
optional reward shaping.

\paragraph{Base two-agent wrapper (common interface).}
The base wrapper constructs an Overcooked-AI environment from a specified layout name and a
fixed finite horizon $H=400$ timesteps. At each timestep, the underlying environment state is
featurized using the Overcooked-AI MDP featurization routine and flattened into a vector
observation
\begin{equation}
o_t = \phi(s_t)\in\mathbb{R}^d.
\end{equation}
The wrapper exposes:
\begin{itemize}
    \item a continuous vector observation space over $\mathbb{R}^d$,
    \item a true two-agent joint action represented as a two-dimensional discrete vector,
    \begin{equation}
    a_t = [a_t^1, a_t^2]\in\{0,\dots,|\mathcal{A}|-1\}^2,
    \end{equation}
    \item the shared team reward returned directly by the underlying Overcooked-AI transition.
\end{itemize}
Each discrete action index $a_t^i$ is mapped to the corresponding primitive Overcooked action,
and the resulting joint action is applied to the simulator. The wrapper then returns the next
flattened observation, the scalar shared reward, and termination signals. Random seeds are
applied consistently by seeding the relevant random number generators before environment resets.

\paragraph{Noise regime.}
The \textbf{Noise} wrapper perturbs each observation component by adding i.i.d. Gaussian noise
with variance $0.01$:
\begin{equation}
\tilde{o}_t = o_t + \epsilon_t,
\qquad
\epsilon_t \sim \mathcal{N}(0, 0.01 I).
\end{equation}
This modification affects only the observation stream and does not alter the underlying
environment dynamics.

\paragraph{Delay regime.}
The \textbf{Delay} wrapper injects stochastic reward delay effects by subtracting a fixed
penalty of $0.5$ from the shared reward with probability $0.2$ at each timestep:
\begin{equation}
\tilde{r}_t =
\begin{cases}
r_t - 0.5, & \text{with probability } 0.2,\\
r_t, & \text{with probability } 0.8.
\end{cases}
\end{equation}
This creates an exogenous disturbance to the reward signal without modifying the base task
objectives.

\paragraph{Combo regime.}
The \textbf{Combo} wrapper applies both perturbations simultaneously, adding i.i.d. Gaussian
noise to observations as above while also applying the probabilistic delay penalty to rewards.

\subsection{Hand-Designed Shaping Baseline (HARL)}
\label{subsec:harl_appendix}

The HARL baseline extends the base wrapper with a simple hand-designed shaping rule that depends
on the number of remaining orders (as reported by the environment). Let
$\mathrm{orders\_remaining}_t$ denote this quantity. The shaped reward is
\begin{equation}
r^{\text{HARL}}_t =
r_t +
\begin{cases}
1.0, & \mathrm{orders\_remaining}_t = 0,\\
0.5, & \mathrm{orders\_remaining}_t < 3,\\
0, & \text{otherwise}.
\end{cases}
\end{equation}
No other changes are made to the PPO policy network, observation encoding, or action interface.
Noisy and delayed variants of HARL are defined by composing this shaping rule with the Noise,
Delay, and Combo wrappers described in Appendix~\ref{subsec:wrappers}.

\subsection{LLM-Based Symbolic Reward Shaping (PPO+LLM)}
\label{subsec:llm_appendix}

The PPO+LLM baseline incorporates a frozen language model as an external evaluator that produces
a binary symbolic assessment of the selected joint action at each timestep. The language model is
EleutherAI GPT-Neo-1.3B and is used strictly in evaluation mode, meaning its parameters are not
updated during reinforcement learning.

\paragraph{Prompt construction and deterministic scoring.}
At each environment step, the wrapper converts the two primitive action indices into a compact
symbolic representation (the standard Overcooked action character encoding) and inserts them into
a fixed natural-language query that asks whether the two joint actions are helpful for cooperative
cooking, with the evaluator constrained to answer using only the words \emph{good} or \emph{bad}.
Instead of sampling a free-form completion, the wrapper performs a deterministic comparison of the
language model scores assigned to the two candidate next-token responses, \emph{good} and \emph{bad}.
Let $\ell_{\text{good}}$ and $\ell_{\text{bad}}$ denote these scores. The symbolic classifier is
\begin{equation}
\mathrm{is\_good}(o_t, a_t) = \mathbb{I}\{\ell_{\text{good}} > \ell_{\text{bad}}\}.
\end{equation}
If the evaluator cannot be queried, the mechanism defaults to a negative assessment, thereby
applying no shaping bonus.

\paragraph{Shaping rule and magnitude.}
If the LLM evaluation is positive, a small constant shaping bonus $\lambda_{s,\mathrm{bonus}}$
is added to the shared reward:
\begin{equation}
r^{\text{LLM}}_t = r_t + \lambda_{s,\mathrm{bonus}}\cdot \mathrm{is\_good}(o_t, a_t),
\end{equation}
with $\lambda_{s,\mathrm{bonus}} = 0.2$ in our experiments. No other aspects of PPO are changed,
and the language model does not modify the PPO policy network architecture.

Noisy and delayed PPO+LLM variants are obtained by composing the same symbolic shaping mechanism
with the Noise, Delay, and Combo wrappers described in Appendix~\ref{subsec:wrappers}.

\subsection{Mapping of Baselines to System Components}
\label{subsec:mapping_appendix}

All baselines share the same trainable PPO backbone described in
Appendix~\ref{subsec:shared_ppo_arch}. They differ only in the environment-side wrapper stack:

\begin{itemize}
    \item \textbf{Baseline, CC\_PPO:} base wrapper under the selected regime (No Noise, Noise, Delay, Combo).
    \item \textbf{SP\_PPO:} identical environment interface to the baseline wrapper, with the difference being
    the training protocol label (self-play setting) rather than a change in policy architecture.
    \item \textbf{HARL:} base wrapper plus the order-based shaping rule in Appendix~\ref{subsec:harl_appendix},
    optionally composed with Noise, Delay, or Combo.
    \item \textbf{PPO+LLM:} base wrapper plus symbolic shaping via frozen GPT-Neo-1.3B as described in
    Appendix~\ref{subsec:llm_appendix}, optionally composed with Noise, Delay, or Combo.
\end{itemize}

\subsection{Architectural Fairness, Parameterization, and Complexity}
\label{subsec:fairness_appendix}

Because all baselines use the same MLP actor-critic backbone and identical PPO hyperparameters,
the number of trainable parameters and representational capacity of the policy and value networks
are constant across all conditions. Consequently, observed performance differences cannot be
attributed to a larger or deeper policy network, but rather to differences in the \emph{system}
induced by wrapper-side perturbations and shaping.

The only additional parameters introduced in PPO+LLM come from the frozen GPT-Neo-1.3B evaluator.
These parameters are not trained and therefore do not increase the expressive capacity of the
PPO policy that selects actions. They do, however, increase per-step inference cost inside the
environment wrapper because each step requires a forward pass through the frozen language model
to compare the scores of the candidate responses \emph{good} and \emph{bad}.

Overall, architectural fairness is enforced by fixing the PPO backbone and varying only the
environment-side structure, namely observation noise, stochastic reward delay, hand-designed
shaping, or LLM-based symbolic shaping.

\subsection{Nash Probe Environment Wrappers and Self-Play Evaluation}
\label{subsec:nash_selfplay_appendix}

In addition to reporting final training returns, we evaluate stability properties of trained
joint policies using a Nash-style probe that compares self-play performance against an approximate
best response. To ensure that this probe is consistent with training, all evaluations reuse the
same wrapper family and environment regimes defined in Appendix~\ref{subsec:wrappers}.

\paragraph{Layout, horizon, and observation encoding.}
All Nash-probe evaluations use the same Overcooked-AI layout employed throughout the paper,
``asymmetric\_advantages'', with the same fixed finite horizon $H=400$.
This layout name refers to a fixed Overcooked-AI kitchen configuration whose geometry and initial
conditions induce asymmetric short-horizon opportunities across chefs, encouraging role
specialization and coordination under shared reward.
Observations are the same flattened featurized vectors $o_t=\phi(s_t)\in\mathbb{R}^d$, and actions
remain the two-agent joint discrete vector $a_t=[a_t^1,a_t^2]$ over the primitive action set
$\mathcal{A}$.

\paragraph{Environment regimes (No Noise, Noise, Delay, Combo).}
For each environment regime $e\in\{\text{No Noise}, \text{Noise}, \text{Delay}, \text{Combo}\}$,
the corresponding wrapper stack is applied exactly as described in Appendix~\ref{subsec:wrappers}.
This guarantees that the Nash probe measures exploitability under the same perturbations that
were present during training for that regime.

\paragraph{Self-play evaluation protocol.}
For each trained joint policy $\pi_{b,e,s}$ (baseline $b$, regime $e$, seed $s$), we estimate
self-play performance by running $N_{\text{eval}}=20$ rollouts in the matching evaluation
environment. Actions are selected deterministically to reduce evaluation variance and isolate
policy differences rather than sampling noise. For rollout $i$, the episodic return is
\begin{equation}
G^{(i)} = \sum_{t=0}^{T^{(i)}-1} r_t^{(i)},
\end{equation}
where $T^{(i)}\leq H$ is the episode length before termination or truncation. The reported
self-play value and standard deviation are
\begin{align}
V_{\text{self}}^{b,e,s} &= \frac{1}{N_{\text{eval}}}\sum_{i=1}^{N_{\text{eval}}} G^{(i)}, \label{eq:self-value}\\
\sigma_{\text{self}}^{b,e,s} &= \sqrt{\frac{1}{N_{\text{eval}}}\sum_{i=1}^{N_{\text{eval}}}\big(G^{(i)}-V_{\text{self}}^{b,e,s}\big)^2}.
\end{align}
All evaluation environments record episode returns and lengths using a consistent logging
procedure. Randomness is controlled by applying the same global seeding procedure used elsewhere
before environment resets.

\subsection{Nash Probe Best-Response Training Procedure}
\label{subsec:nash_br_train_appendix}

To measure unilateral exploitability of each joint policy $\pi_{b,e,s}$, we train an approximate
best response against it under a fixed compute budget, using PPO with the same trainable backbone
described in Appendix~\ref{subsec:shared_ppo_arch}. Concretely, for each $(b,e,s)$, we construct a
single-agent training problem in which one agent is optimized while its partner is held fixed.

\paragraph{Induced single-agent best-response MDP.}
Given a trained joint policy $\pi_{b,e,s}$, we define an induced single-agent MDP
$M_{\text{BR}}(\pi_{b,e,s})$ as follows. The environment state and observation encoding are identical
to the two-agent setting (same layout, same wrapper regime $e$, same featurization and flattening),
but the learning agent controls only one player index $i$. In our experiments we fix $i=0$ to probe
unilateral deviations of the first agent, while the partner agent follows the stored policy.
At each timestep:
\begin{itemize}
    \item the learning agent selects a single primitive action index $a_t^{\text{BR}}\in\{0,\dots,|\mathcal{A}|-1\}$,
    \item the frozen partner action is obtained by a deterministic evaluation of the stored policy,
          yielding $\hat{a}_t^{1-i}$,
    \item the joint action applied to Overcooked is $a_t = [a_t^0,a_t^1]$ where the $i$th component is
          replaced by $a_t^{\text{BR}}$ and the other component is $\hat{a}_t^{1-i}$.
\end{itemize}
The environment then advances one step and returns the next observation, the shared team reward,
and termination signals to the learning agent. The learning agent therefore optimizes expected team
return while its partner follows the fixed self-play policy.

\paragraph{Best-response training setup.}
For each policy $\pi_{b,e,s}$ we train one best response using PPO for a fixed budget of
\begin{equation}
T_{\text{BR}} = 200{,}000
\end{equation}
environment timesteps. PPO uses the same MLP actor-critic backbone and the same core
hyperparameters as in joint-policy training (rollout length $2048$, batch size $2048$, learning rate
$3\times 10^{-4}$, and discount factor $\gamma=0.99$). To decouple best-response randomness from the
original training run while keeping the process reproducible, the best-response random seed is set to
\begin{equation}
s_{\text{BR}} = s + 999,
\end{equation}
and the same global seeding routine is applied prior to environment resets and PPO initialization.

\paragraph{Best-response evaluation.}
After training, we evaluate the learned best-response policy for $N_{\text{eval}}=20$ rollouts in the
same best-response environment, again using deterministic action selection. Let $\tilde{G}^{(i)}$
denote the episodic return when the learning agent follows the trained best-response policy and the
partner follows the frozen policy $\pi_{b,e,s}$. We report
\begin{align}
V_{\text{BR}}^{b,e,s} &= \frac{1}{N_{\text{eval}}}\sum_{i=1}^{N_{\text{eval}}} \tilde{G}^{(i)}, \label{eq:br-value}\\
\sigma_{\text{BR}}^{b,e,s} &= \sqrt{\frac{1}{N_{\text{eval}}}\sum_{i=1}^{N_{\text{eval}}}\big(\tilde{G}^{(i)}-V_{\text{BR}}^{b,e,s}\big)^2}.
\end{align}
These quantities are used in the main text to compute the Nash-gap statistic
$\Delta^{b,e,s}=V_{\text{BR}}^{b,e,s}-V_{\text{self}}^{b,e,s}$, which measures how much a unilateral
deviation (within the chosen PPO best-response class and training budget) can improve expected team
return under the same evaluation regime.

\subsection{Parallel Execution and Resume Logic for the Nash Probe}
\label{subsec:nash_parallel_appendix}

The Nash-probe analysis evaluates all combinations of baselines, environment regimes, and random
seeds, namely the Cartesian product
\begin{equation}
\mathcal{J} = \mathcal{B}\times\mathcal{E}\times\mathcal{S}_{\text{train}},
\end{equation}
subject to the practical constraint that a corresponding trained joint-policy checkpoint exists.
Each job $(b,e,s)\in\mathcal{J}$ produces a single result tuple consisting of self-play statistics,
best-response statistics, and their difference (the Nash gap).

\paragraph{Crash-safe result logging.}
To make the computation resumable, the analysis maintains a persistent results log stored as a
comma-separated table with one row per completed triple $(b,e,s)$. Each row records the baseline
identifier, environment regime, seed, the self-play mean return, the self-play return standard
deviation, the best-response mean return, the best-response return standard deviation, and the
resulting Nash gap. Appending results as each job completes ensures that partial progress is
preserved across interruptions.

\paragraph{Skipping completed jobs on restart.}
At startup, the analysis reads the persistent results log (if it exists) and collects the set of
already-processed triplets $(b,e,s)$. The remaining job list is then constructed as the subset of
$\mathcal{J}$ that is not yet recorded. This guarantees that each triple is evaluated at most once
and that rerunning the analysis resumes from the last completed entry without duplicating work.

\paragraph{One-time environment initialization for robust parallelism.}
Before launching parallel workers, the analysis performs a one-time initialization pass that
constructs an Overcooked gridworld instance and runs the featurization procedure on an initial
state. This initializes internal Overcooked structures, including the medium-level action manager,
in the main process so that their internal data are created once and can be safely reused by worker
processes. This reduces repeated initialization cost and avoids common serialization issues that can
arise when workers attempt to independently construct complex environment objects.

\paragraph{Parallel job execution.}
The remaining jobs are processed in parallel using process-based execution with six worker
processes. Each worker executes the same routine for its assigned $(b,e,s)$:
\begin{itemize}
    \item It checks whether the required trained joint-policy checkpoint for $(b,e,s)$ exists. If not,
          the job produces no result and is skipped.
    \item Otherwise, it evaluates self-play performance for $N_{\text{eval}}=20$ episodes in the matching
          regime $e$, computing the mean and standard deviation of episodic return.
    \item It then trains an approximate best response in the induced best-response environment for a fixed
          budget of $T_{\text{BR}}=200{,}000$ timesteps, using the same PPO backbone and hyperparameters as
          described earlier in this appendix.
    \item Finally, it evaluates the trained best response for $N_{\text{eval}}=20$ episodes and computes the
          Nash gap $\Delta^{b,e,s}=V_{\text{BR}}^{b,e,s}-V_{\text{self}}^{b,e,s}$.
\end{itemize}
After completing a job, the worker appends exactly one new row to the persistent results log. This
append-only design ensures that partial progress is preserved, enables safe restarts, and supports
long-running sweeps over $\mathcal{J}$ without requiring all results to be held in memory.

This design ensures that (i) each triple $(b,e,s)$ is evaluated at most once, (ii) partial progress
is preserved across runs via the persistent results log, and (iii) the Nash probe remains consistent
with the original training configuration by reusing the same layout, horizon, observation encoding,
wrapper regimes, and PPO hyperparameters described in this appendix.

\makeatletter
\eoddefinedtrue
\makeatother

\end{document}